\documentclass[12pt,preprint]{aastex} 

\slugcomment{Accepted for publication in ApJ}

\shorttitle{Optical spectroscopy of XUV disks}
\shortauthors{Gil de Paz et al.}

\begin{document}

\title{Chemical and Photometric Evolution of Extended Ultraviolet Disks: Optical Spectroscopy of M\,83 (NGC~5236) and NGC~4625\altaffilmark{1}}

\author{Armando Gil de Paz\altaffilmark{2,3}, Barry F. Madore\altaffilmark{2,4}, Samuel Boissier\altaffilmark{2,5}, David Thilker\altaffilmark{6}, Luciana Bianchi\altaffilmark{6}, Carmen S\'{a}nchez Contreras\altaffilmark{7,8}, Tom A. Barlow\altaffilmark{8}, Tim Conrow\altaffilmark{8}, Karl Forster\altaffilmark{8}, Peter G. Friedman\altaffilmark{8}, D. Christopher Martin\altaffilmark{8}, Patrick Morrissey\altaffilmark{8}, Susan G. Neff\altaffilmark{9}, R. Michael Rich\altaffilmark{10}, David Schiminovich\altaffilmark{8}, Mark Seibert\altaffilmark{8}, Todd Small\altaffilmark{8}, Jos\'{e} Donas\altaffilmark{4}, Timothy M. Heckman\altaffilmark{11}, Young-Wook Lee\altaffilmark{12}, Bruno Milliard\altaffilmark{4}, Alex S. Szalay\altaffilmark{11}, Ted K. Wyder\altaffilmark{8}, Sukyoung Yi\altaffilmark{12}}

\altaffiltext{1}{Based in part on observations made at Magellan~I (Baade) telescope which is operated by the Carnegie Institution of Washington.}
\altaffiltext{2}{The Observatories, Carnegie Institution of Washington, 813 Santa Barbara Street, Pasadena, CA 91101; agpaz@ociw.edu}
\altaffiltext{3}{Departamento de Astrof\'{\i}sica, Universidad Complutense de Madrid, Madrid 28040, Spain; agpaz@astrax.fis.ucm.es}
\altaffiltext{4}{NASA/IPAC Extragalactic Database, California Institute of Technology, MS 100-22, Pasadena, CA 91125; barry@ipac.caltech.edu} 
\altaffiltext{5}{Laboratoire d'Astrophysique de Marseille, BP 8, Traverse du Siphon, 13376 Marseille Cedex 12, France; samuel.boissier, jose.donas, bruno.milliard@oamp.fr}
\altaffiltext{6}{Center for Astrophysical Sciences, The Johns Hopkins University, 3400 N. Charles St., Baltimore, MD 21218; dthilker, bianchi@pha.jhu.edu}
\altaffiltext{7}{Instituto de Estructura de la Materia, CSIC, Serrano 121, Madrid 29006, Spain; carmen@damir.iem.csic.es}
\altaffiltext{8}{California Institute of Technology, MC 405-47, 1200 East California Boulevard, Pasadena, CA 91125; tab, tim, krl, friedman, cmartin, patrick, ds, mseibert, tas, wyder@srl.caltech.edu}
\altaffiltext{9}{Laboratory for Astronomy and Solar Physics, NASA Goddard Space Flight Center, Greenbelt, MD 20771; neff@stars.gsfc.nasa.gov}
\altaffiltext{10}{Department of Physics and Astronomy, University of California, Los Angeles, CA 90095; rmr@astro.ucla.edu}
\altaffiltext{11}{Department of Physics and Astronomy, The Johns Hopkins University, Homewood Campus, Baltimore, MD 21218; heckman, szalay@pha.jhu.edu}
\altaffiltext{12}{Center for Space Astrophysics, Yonsei University, Seoul 120-749, Korea;  ywlee@csa.yonsei.ac.kr, yi@astro.ox.ac.uk}

\begin{abstract}
We present the results from the analysis of optical spectra of 31
H$\alpha$-selected regions in the extended UV (XUV) disks of M\,83
(NGC~5236) and NGC~4625 recently discovered by GALEX. The spectra were
obtained using IMACS at Las Campanas Observatory 6.5m Magellan~I
telescope and COSMIC at the Palomar 200-inch telescope, respectively
for M\,83 and NGC~4625. The line ratios measured indicate nebular
oxygen abundances (derived from the R23 parameter) of the order of
Z$_{\odot}$/5-Z$_{\odot}$/10\footnote{We adopt
12+log(O/H)$_{\odot}$=8.69 (Allende Prieto, Lambert, \& Asplund 2001)
throughout the paper.}. For most emission-line regions analyzed the
line fluxes and ratios measured are best reproduced by models of
photoionization by single stars with masses in the range
20-40\,M$_{\odot}$ and oxygen abundances comparable to those derived
from the R23 parameter. We find indications for a relatively high N/O
abundance ratio in the XUV disk of M\,83. Although the metallicities
derived imply that these are not the first stars formed in the XUV
disks, such a level of enrichment could be reached in young spiral
disks only 1\,Gyr after these first stars would have formed. The
amount of gas in the XUV disks allow maintaining the current level of
star formation for at least a few Gyr.
\end{abstract}


\section{Introduction}
\label{introduction}

Studying the outer edges of spiral galaxies provides a unique tool for
understanding the formation and evolution of galactic disks. According
to the {\it inside-out} scenario of disk formation, an increase in the
gas-infall timescale with the galactocentric radius results in the
delayed formation of the stars in the outer parts of the disk compared
with its inner regions (Larson 1976, Matteucci \& Francois 1989, Hou
et al$.$ 2000, Prantzos \& Boissier 2000). This scenario, originally
proposed to explain the color and metallicity gradients in the Milky
Way, is also supported by recent N-body/SPH simulations of the
evolution of galactic disks (e.g$.$ Brook et al$.$ 2006).

Thus, star formation taking place now in some of the outermost regions
of galaxies might probe physical conditions similar to those present
during the formation of the first stars in the Universe. The low gas
densities found in these regions also provide an excellent test for
the presence (or absence) of a threshold for star formation (Martin \&
Kennicutt 2001) and for studying the behavior of the star-formation
law in low-density rarefied environments (Kennicutt 1989, Boissier et
al$.$ 2003, 2006, Elmegreen \& Hunter 2006).

Recent deep, wide-field observations of a sample of nearby spiral
galaxies at UV wavelengths carried out by the Galaxy Evolution
Explorer (GALEX) satellite as part of its Nearby Galaxies Survey
(Bianchi et al$.$ 2003; see also Gil de Paz et al$.$ 2006) have
revealed the presence of UV-bright complexes (XUV complexes hereafter)
in the outermost parts of their disks. These galaxies, referred to as
XUV-disk galaxies, host UV emission well beyond ($\sim$2-4$\times$)
their optical (D25) radii (Thilker et al$.$ 2005a, 2007, in prep$.$,
Gil de Paz et al$.$ 2005). To date, the two best-studied XUV disks are
those of M\,83 (Thilker et al$.$ 2005a; D=4.5\,Mpc) and NGC~4625 (Gil
de Paz et al$.$ 2005; D=9.5\,Mpc). These are two quite different
examples of XUV disks: M\,83 has a very massive, large,
high-surface-brightness optical disk with patchy XUV emission clearly
disconnected from it; NGC~4625 is a low-luminosity system with
ubiquitous XUV emission on top of an underlying low-surface-brightness
optical disk (see Swaters \& Balcells 2002).

Any study of the properties of the XUV complexes discovered in M\,83
and NGC~4625 must establish whether the UV emission associated with
these extended disks is due to recent star formation or not. If that
is the case, as previous studies have suggested, a detailed analysis
of their properites should then provide fundamental clues to
understand the possible mechanism(s) that led to formation of stars in
these outermost regions of the disks. These can be internal
mechanisms, such as density waves, or external influences, such as
tidal interactions or satellite disruption. Finally, we should be able
to establish whether (1) this is continously happening in the outer
disks of these galaxies, (2) it is a transient but recurrent
phenomenon that has taken place in these and other (perhaps all)
spiral galaxies in the past, or, alternatively, (3) the XUV emission
is a one-time phenomenon and these are the first generation of stars
to form in the outer parts of these galaxies.

In order to gain some deeper insight into the properties of these XUV
complexes we have obtained deep optical spectroscopy of the outer disk
of M\,83 and NGC~4625 using the Magellan and Palomar 200-inch
telescopes, respectively. In this work we study the physical
conditions of the ionized gas (densities, metal abundances) in those
XUV complexes showing H$\alpha$ emission. We also estimate the
properties of the ionizing sources by comparing the line ratios
measured with the predictions of photionization models; in particular,
we determine if these line ratios are consistent with the line
emission being due to photoionization by single stars as has been
recently suggested (Gil de Paz et al$.$ 2005).

The spectroscopic observations of the XUV disks of M\,83 and NGC~4625
along with the data reduction methodology are described in
Section~\ref{observations}. In that section we also briefly describe
complementary optical and GALEX UV imaging data. Section~\ref{results}
presents the results from the analysis of the spectroscopic data. We
discuss the nature of the XUV emission and its possible implications
for the past and future evolution of these systems in
Section~\ref{discussion}. The conclusions are summarized in
Section~\ref{conclusions}.

\section{Observations and reduction}
\label{observations}

\subsection{GALEX UV imaging}

The regions analyzed here were originally identified in UV images
taken by the GALEX satellite (Thilker et al$.$ 2005a, Gil de Paz et
al$.$ 2005). The Galaxy Evolution Explorer (GALEX) satellite is a
NASA's small explorer mission with a single instrument which consists
of a Ritchey-Chr\'{e}tien telescope with a 50-cm aperture that allows
simultaneous imaging in two UV bands/channels, FUV
($\lambda_{\mathrm{eff}}$=1516\,\AA) and NUV
($\lambda_{\mathrm{eff}}$=2267\,\AA), within a circular field of view
of 1.2\,degrees in diameter. See Martin et al$.$ (2005) for a more
detailed description of the mission and the GALEX instrument.

The GALEX observations used in this paper were carried out on 2003
June 7 in the case of M\,83 and on 2004 April 5 for NGC~4625. The total
exposure times (equal in both the FUV and NUV bands) were 1352\,s and
1629\,s for M\,83 and NGC~4625, respectively. These data were all
reduced using the standard GALEX pipeline. 

\subsection{Optical imaging}

Ground-based optical imaging has been carried out to obtain accurate
positions for the regions responsible for the XUV emission. Accuracies
of the order of a few tenths of arc second are needed for obtaining
multi-object optical spectroscopy with relatively good spectral
resolution for the 1-1.5\,arcsec-wide slitlets used (see
Section~\ref{observations.spectroscopy}).

On 2005 March 10-14, we obtained deep broad-band $UR$ imaging data of
the southern part of the XUV disk of M\,83 using the
2048$\times$3150\,pixels Site~3 Direct CCD at the Cassegrain focus of
the Swope 1-m telescope at Las Campanas Observatory (Chile) (see
Figure~1). The exposure time was 6$\times$1800\,seconds in $U$ and
3$\times$900\,seconds in $R$. Deep ground-based optical imaging of
NGC~4625 in $B$ and $R$ bands had been previously obtained at the
Isaac Newton 2.5-m telescope in La Palma (Spain) on 1995 May 28 and
1995 Dec 24 \& 27 (see Figure~2). Images are available through the
Isaac Newton Group (ING) data archive. See also Gil de Paz et al$.$
(2005) and Swaters \& Balcells (2002) for details on the optical
imaging observations of NGC~4625 (see also Table~1).

Many of the galaxies with extended UV-bright disks found by GALEX so
far show a relatively sharp cutoff in the azimuthally-averaged
surface-brightness profile in H$\alpha$ at the limit of the optical
disk, while the UV light apparently extends smoothly beyond that
radius (Meurer et al$.$ 2004, Thilker et al$.$ 2005a, Gil de Paz et
al$.$ 2005). It is therefore expected that a significant fraction of
the regions identified in the GALEX images would not show any line
emission and, consequently, could not be used for determining the
ionized-gas metal abundances and dust reddening (from measurements of
the Balmer-line decrement). With this in mind and in order to improve
the efficiency of our spectroscopic observations, we also obtained
deep narrow-band H$\alpha$ imaging of both M\,83 and NGC~4625. M\,83
was observed on 2005 March 14 through a 70-\AA-wide narrow-band filter
centered on 6570\AA\ using the 2048$\times$3150\,pixels Site~3 Direct
CCD attached to the Swope 1-m telescope in Las Campanas. The total
exposure time in H$\alpha$ was 10,200\,seconds split in 9
exposures. On 2004 August 20, we obtained an 800 second narrow-band
image of NGC~4625 through a 20-\AA-wide filter centered on 6563\AA\
using COSMIC mounted at the prime focus of the Palomar Observatory 5-m
Hale telescope (Kells et al$.$ 1998).

Reduction of the images was carried out using IRAF\footnote{IRAF is
distributed by the National Optical Astronomy Observatories, which are
operated by the Association of Universities for Research in Astronomy,
Inc., under cooperative agreement with the National Science
Foundation.}. For the H$\alpha$ imaging data, the continuum was
subtracted using a scaled $R$-band image such that the fluxes of the
field stars in both images were equal. The astrometry of the optical
images was carried out using the IRAF tasks {\sc starfind}, {\sc
ccxymatch}, {\sc ccmap}, and {\sc ccsetwcs}. We made use of the
USNO-B1 catalog in the case of M\,83, and the USNO-A2 catalog for the
images of NGC~4625. The r.m.s$.$ achieved by our astrometric
calibration was smaller than 0.3\,arcsec in all cases. The procedures
followed for subtracting the continuum from the H$\alpha$ images and
for performing the astrometry of the optical images are described in
detail in Gil de Paz, Madore, \& Pevunova (2003).

False-color RGB composites using GALEX FUV, broad-, and narrow-band
ground-based optical images are shown in Figure~1 for M\,83 and in
Figure~2 for NGC~4625. A summary of the UV and optical imaging
observations is given in Table~1.

\subsection{Optical spectroscopy}
\label{observations.spectroscopy}

\subsubsection{Sample selection}
\label{observations.sample}

Our spectroscopic sample is made up primarily of regions with detected
emission in the continuum-subtracted narrow-band images. The vast
majority of these regions correspond to stellar associations that also
show bright UV emission. Additionally, we included $U$-bright sources
with no obvious H$\alpha$ counterpart, mainly to take advantage of the
large field-of-view offered by the IMACS spectroscopic observations of
M\,83. In this paper we will focus exclusively on the analysis of the
properties of the emission-line regions.

In Table~2 we give the coordinates of the emission-line regions in the
XUV disks of M\,83 and NGC~4625. These are all the regions for which
our follow-up spectroscopic observations confirmed both their nature
as emission-line sources (both H$\alpha$ and H$\beta$ being detected
in emission) and having a redshift close to that of the corresponding
parent galaxy. The coordinates given in this table are accurate to the
level given by the r.m.s$.$ of our astrometric calibration, i.e$.$
better than 0.3\,arcsec. Small systematic errors inherent to the USNO
catalog could be also present.

\subsubsection{M\,83 observations}

Multi-object spectroscopic observations of the southern region of the
XUV disk of M\,83 (see Figure~1) were obtained using the IMACS
spectrograph at the 6.5-m Magellan-I (Baade) telescope in Las Campanas
(Chile). The observations were carried out on 2005 April 18 using the
f/4 long camera with the 300\,l\,mm$^{-1}$ grating in its first
order. This configuration provides a spatial scale of
0.11\,arcsec\,pixel$^{-1}$, a reciprocal dispersion of
0.743\AA\,pixel$^{-1}$, and a spectral resolution of R$\sim$1200 for
slitlets of 1\,arcsec in width. IMACS uses a 8K$\times$8K-pixel mosaic
camera. The objects observed were distributed over two different masks
with total exposure times of 40 and 60\,minutes. During the night the
seeing ranged between 0.6 to 1.0\,arcsec.

The spectra obtained were reduced using the so-called COSMOS package,
an IMACS data reduction pipeline developed by August Oemler. COSMOS
consists of a set of tasks that allow correcting the science images
for bias and flat-fielding, subtracting the sky, and performing a
precise mapping of the CCD pixels onto the coordinate system of
wavelength and slit position using the observing setup and mask
generation files as input.

In the case of the IMACS observations of M\,83 a total of 37
H$\alpha$-selected (34 in the XUV disk) and 32 $U$-band-selected
sources were included in the two masks observed. Out of the 37
H$\alpha$-selected sources 22 of them (19 out of 34 in the XUV disk)
were detected both in the H$\alpha$ and H$\beta$ lines in our
spectroscopic data with redshifts close to that of M\,83. The
positions and observed H$\beta$ fluxes inside the slitlets for these
22 regions are given in Table~3. We also serendipitously discovered a
background galaxy ([GMB2007]1) at $z$$\simeq$0.3
[RA(J2000)=13$^{\mathrm{h}}$37$^{\mathrm{m}}$24.53$^{\mathrm{s}}$;
Dec(J2000)=$-$29$^{\circ}$57'20.1''] whose emission in
[OIII]$\lambda$5007\AA\ falls in the window defined by the filter used
in our narrow-band imaging observations and for which we also detected
H$\beta$ and [OIII]$\lambda$4959\AA.

\subsubsection{NGC~4625 observations}

The spectroscopic observations of the XUV disk of NGC~4625 were
carried out using COSMIC mounted at the prime focus of the Palomar
Observatory 5-m (Hale) telescope. The spectra were taken on the
night of 2005 March 16 using the 300\,l\,mm$^{-1}$ grating and a
mask with 1.5\,arcsec-wide slitlets. COSMIC uses a single
2048$\times$2048-pixels CCD detector. The total exposure time of the
single mask observed was 40\,minutes and the seeing during the
observations was $\sim$1.2\,arcsec. The spectra were reduced using
standard IRAF tasks within the {\sc ccdred}, {\sc onedspec}, and {\sc
twodspec} packages.

Out of the 17 H$\alpha$-selected targets included in the mask, 12
sources showed emission in at least H$\alpha$ and H$\beta$ at the
redshift of NGC~4625 and were used in our analysis (see Table~3 for
the list of positions and observed fluxes in H$\beta$). Additionally,
we identified a background galaxy ([GMB2007]2) at $z$$\simeq$0.3
located at
RA(J2000)=12$^{\mathrm{h}}$41$^{\mathrm{m}}$41.89$^{\mathrm{s}}$ and
Dec(J2000)=+41$^{\circ}$16'10.9''.

\subsubsection{Flux calibration}
\label{calibration}

The spectroscopic observations of M\,83 and NGC~4625 were both made
under photometric conditions. In the case of M\,83 we observed the
spectroscopic standard star LTT~3218 using a 2.5\,arcsec-wide
long-slit mask. We adopted the extinction curve for Cerro Tololo
Inter-American Observatory. For the observations of NGC~4625 we
obtained spectra of BD$+$33$^{\circ}$2642 and HD~93521 using a
4\,arcsec-wide long-slit mask. The extinction curve of Kitt Peak
National Observatory was adopted in this case.

The flux-calibrated spectra obtained for the 22 (12)
H$\alpha$-selected regions in M\,83 (NGC~4625) used for this work are
shown in Figure~3 (Figure~4). Since the observation of the
spectrophotometric standard stars was performed near the center of the
field we compared the resulting flux-calibrated sky spectra obtained
through each of the slitlets in order to ensure the validity of this
calibration for all sources in the mask. 

For M\,83 the absolute calibration shows a dispersion of
$\sim$0.3\,mag (Figure~5a). However, if we apply systematic offsets to
the individual spectra (which do not affect the line ratios measured)
the dispersion between these spectra due to changes in the wavelength
dependence of the flux calibration is reduced to $\sim$0.15\,mag
between 4000-6000\,\AA\ and $\sim$0.2\,mag below 4000\,\AA\ and above
6000\,\AA\ (Figure~5b). These numbers represent the average 1-$\sigma$
calibration errors for our spectra.

In the case of NGC~4625 both the absolute and relative differences
between all the sky spectra extracted are significanly smaller than
for M\,83 (see Figures~5c and 5d). Once small global offsets are
applied the r.m.s$.$ between the spectra becomes $\sim$0.1\,mag for
$\lambda$$<$4300\,\AA\ or $\lambda$$>$6600\,\AA\ or even smaller for
4300\,\AA$<$$\lambda$$<$6600\,\AA. This is probably due to the much
reduced field of view of this instrument and smaller format of the
detector.

\section{Results}
\label{results}

\subsection{Line fluxes and ratios}
\label{results.lineratios}

The extinction-corrected H$\beta$ line fluxes measured through our
slitlets are given in Table~2. These spectroscopic fluxes all below
the value expected for a region ionized by a single O3 star
(L$_{\mathrm{H}\alpha}$$\simeq$10$^{38}$\,erg\,s$^{-1}$; Sternberg et
al$.$ 2003) at the distance of our targets. Similar conclusions were
reached by Gil de Paz et al$.$ (2005) through the analysis of
narrow-band imaging data on NGC~4625. Note that although many of these
regions are very compact, aperture effects consequence of the limited
size of our slitlets (1-1.5\,arcsec in width) might be
significant. Therefore, the spectroscopic line fluxes given in Table~2
should be considered as lower limits to the total observed flux of the
corresponding HII region.

In Table~3 we present the line ratios measured for each of the 19 (12)
confirmed line-emitting regions in the XUV disk of M\,83 (NGC~4625),
plus three additional regions in its optical disk (NGC~5236: XUV 20,
21, and 22). Both fluxes and line ratios have been corrected for
foreground Galactic and internal extinction using the
H$\alpha$/H$\beta$ Balmer-line decrement. We have adopted an intrinsic
value for H$\alpha$/H$\beta$ of 2.86 for case B recombination
(Osterbrock 1989) and the parametrization of the Galactic extinction
law given by Cardelli, Clayton, \& Mathis (1989).

Figure~6 shows the standard diagnostic diagrams proposed by Baldwin,
Phillips, \& Terlevich (1981) and Veilleux \& Osterbrock (1987). The
line ratios measured in the XUV disk of M\,83 (panels~6a \& 6b) are
similar to those found in local star-forming galaxies (SFG),
represented here by the H$\alpha$-selected UCM-Survey galaxies of
Zamorano et al$.$ (1994, 1996). We find that regions in the inner
(optical) disk of M\,83 ($r$$<$5.5'; filled circles) have lower
excitations than those in the inner part of the XUV disk
(5.5$<$$r$$<$10'; filled triangles) and even lower than in the
outermost XUV disk ($r$$>$10'; filled stars).

On other hand, the line-emitting regions in the XUV disk of NGC~4625
show relatively high excitation conditions. We also find a larger
dispersion in the [OIII]$\lambda$5007\AA/H$\beta$ ratio for a given
value of [NII]$\lambda$6584\AA/H$\alpha$ that in the case of
M\,83. Gil de Paz et al$.$ (2005) showed that many of the regions
identified in the XUV disk of NGC~4625 have H$\alpha$ luminosities
consistent with photoionization due to single massive
stars. Interestingly, simultaneous low [OIII]$\lambda$5007\AA/H$\beta$
and [NII]$\lambda$6584\AA/H$\alpha$ ratios are predicted by some of
the single-star photoionization models of Stasinska \& Schaerer
(1997). These models do not cover the entire parameter space of the
properties of the ionizing star (luminosity, metallicity, etc$.$) or
the surrounding HII region (density, geometry) nor were intended for
reproducing the line ratios of single-star HII regions but their
ionization structure. Thus, a detailed modelling of all the observed
line ratios for each individual region covering a wider range of
parameters than those explored by Stasinska \& Schaerer (1997) is
carried out in Section~\ref{cloudy} (see also Figure~6).

Panels~6b and 6d show the predictions of the photoionization models
for evolving starbursts developed by Stasinska \& Leitherer (1996). In
Figure~6b the models represent the variation in the line ratios with
age between 1 to 10\,Myr for different metallicites (Z$_{\odot}$,
Z$_{\odot}$/4, Z$_{\odot}$/10, Z$_{\odot}$/40). For starburst models
with metallicities below solar the sequence in excitation conditions
seen in these figures is mainly due to a change in age, from young
high-excitation starbursts (whose effective temperature is dominated
by very massive stars) at the top-left part of the diagram to more
evolved low-excitation starbursts that can be found at the
bottom-right.

The fact that the loci of the emission-line XUV complexes in these
diagrams coincide with those of local star-forming galaxies indicates
that the line emission detected is produced by photoionization by
young massive stars and not by planetary nebulae, which typically show
lower [OII]$\lambda\lambda$3727\AA\AA/H$\beta$ and somewhat higher
[OIII]$\lambda$5007\AA/H$\beta$ line ratios than those found here (see
e.g$.$ Baldwin et al$.$ 1981). The compact nature of these sources
(see Figures~1 \& 2) also excludes the possibility of this line
emission being associated with Diffuse Ionized Gas (DIG).

\subsection{Metal abundances based on strong lines}
\label{abundances}

Determining the metallicity of the ionized gas in star-forming
galaxies and HII regions generally means measuring their oxygen
abundances. Ideally this requires deriving the electron temperature
from the measurement of a temperature-sensitive line ratio such as
[OIII]$\lambda$4363\AA/[OIII]$\lambda$5007\AA. Unfortunately, the
[OIII]$\lambda$4363\AA\ line is very weak and can only be detected in
bright, actively star-forming low-metallicity systems.

Since the [OIII]$\lambda$4363\AA\ line in not detected in our spectra,
in order to constrain the metallicities of the emission-line regions
in the XUV disks of M\,83 and NGC~4625 we rely on spectral diagnostics
based on strong lines such as those described in Kewley \& Dopita
(2002). Among these, perhaps the best estimator of the nebular oxygen
abundance is the R23 ratio. Defined as
log\{([OII]$\lambda\lambda$3727\AA\AA$+$[OIII]$\lambda$5007\AA$+$[OIII]$\lambda$4959\AA)/H$\beta$\},
the R23 parameter was first introduced by Pagel et al$.$
(1979). Although very sensitive to metal abundance, this parameter is
also sensitive to the ionization parameter, especially at low
metallicities (Z$<$Z$_{\odot}$). This degeneracy between metallicity
and ionization parameter can be partially solved by taking into
account the value of an indicator mostly sensitive to the ionization
parameter, such as the
[OII]$\lambda\lambda$3726,3729\AA\AA/[OIII]$\lambda$5007\AA\
([OII]/[OIII] hereafter) ratio (e.g$.$ McGaugh 1991).

Another critical issue regarding the use of the R23 ratio is the fact
that is double-valued for most R23 flux ratios. It shows a maximum
that, depending on the ionization parameter, is located anywhere
between half solar to solar metallicity and below that maximum the
value of the R23 can be reproduced by either a low-metallicity or a
high-metallicity photoionization model. In order to solve this problem
an initial guess on the oxygen abundance based on some other
diagnostics or secondary indicators, such as the
[NII]$\lambda$6584\AA/[OII]$\lambda\lambda$3726,3729\AA\AA\
([NII]/[OII]),
[NII]$\lambda$6584\AA/[SII]$\lambda\lambda$6717,6731\AA\AA\
([NII]/[SII]), or [NII]$\lambda$6584\AA/H$\alpha$ ([NII]/H$\alpha$)
ratios, is desirable.

We have used two different recipes, those published by McGaugh (1991;
M91 hereafter) and Pilyugin \& Thuan (2005; PT05 hereafter), which
allow for a determination of the oxygen abundance as a simple function
of the R23 ratio and the [OII]/[OIII] line ratio, once either the low-
or high-metallicity branch is chosen. Note that for high metallicities
only the method of PT05 includes a dependence on the [OII]/[OIII] line
ratio through the use of the excitation parameter, $P$ \{$\equiv$([OIII]$\lambda$5007\AA$+$[OIII]$\lambda$4959\AA)/([OII]$\lambda\lambda$3727\AA\AA$+$[OIII]$\lambda$5007\AA$+$[OIII]$\lambda$4959\AA)\}.

\subsubsection{M\,83}
\label{strong.m83}

The large dispersion in the line ratios measured across the disk of
the M\,83 indicates a wide range of ionization conditions and/or metal
abundances. Therefore, we should be very cautious in choosing what
branch of the parametrization of the R23 ratio to use.

In the case of the three innermost regions analyzed in this paper
(NGC~5236: XUV 20, 21, and 22), which are still considered part of the
optical disk of the galaxy, the [NII]/[OII] ratios measured are high
enough \{log([NII]/[OII])$>$0 in all three cases\} that we can
confidently assume the high-metallicity-branch for the calibration of
the R23 parameter. For the regions in the XUV disk of M\,83 the
[NII]/[OII] ratios measured (between $-$1.1 and $-$0.3 in log scale)
indicate metal abundances around solar. The [NII]/[SII] ratios
measured, ranging between $-$0.2 and 0.4 in logarithmic scale but with
most of the regions showing positive values, favors again metal
abundances close to solar. In the case of the [NII]/H$\alpha$ line
ratios, whose values range approximately between $-$1 and $-$0.4, we
can only say that metallicities are probably above Z$_{\odot}$/2.  We
should point out that all [NII]/[OII], [NII]/[SII], and
[NII]/H$\alpha$ are good diagnostics for the oxygen abundance as long
as the N/O abundance ratio adopted by Kewley \& Dopita (2002) is
appropriate for these regions (see below for further
discussion). These numbers do not enable us to determine which of the
two branches of the calibration of the R23 ratio with metallicity
should be used in the case of the emission-line regions of the XUV
disk of M\,83. Accordingly, in Table~3 we give the values derived
assuming both the low- and high-metallicity cases for both the M91 and
PT05 calibrations. The fact that these secondary indicators suggest
metal abundances close to the limit between the low- and
high-metallicity branches of the R23 calibration results in very
similar oxygen abundance estimates under the two assumptions for
regions NGC~5236: XUV 07 and NGC~5236: XUV 17.

The R23 calibration of PT05 yields systematically lower oxygen
abundances than those derived using the M91 method (similar systematic
differences have been found by van Zee \& Haynes 2006 when comparing
the calibrations of M91 and Pilyugin 2000). This is the case for both
the lower and upper branch of the calibration. These differences range
between $\leq$0.1\,dex in regions NGC~5236: XUV 01 and 09, NGC~4625:
XUV 01 and 07 to $\geq$0.4\,dex in regions NGC~5236: XUV 07 and 17,
and NGC~4625: XUV 06. The largest differences between the predictions
of these two methods are found in regions located in the so-called
{\it transition zone} (PT05). It is worth mentioning that the lower
[upper] branch of the empirical calibration of PT05 was limited to
oxygen abundances 12+log(O/H)$<$8.00 [12+log(O/H)$>$8.25] and values
of the excitation parameter, 0.55$<$$P$$<$1 [$P$$>$0.1]. Thus, oxygen
abundances in Table~3 not fulfilling these criteria are extrapolated
values and should be considered with caution.

The photoionization models of Kewley \& Dopita (2002) assume a N/O
abundance ratio that is approximately constant
\{log(N/O)$\sim$$-$1.5; i.e$.$ nitrogen is assumed of primary origin\} at low oxygen
abundances \{12+log(O/H)$<$8.5\} and rises linearly with the oxygen
abundance above this value (i.e$.$ secondary nitrogen). They used an
empirical fit to the abundance measurements of van Zee, Haynes
\& Salzer (1998a). However, the N/O ratio is highly dependent on the
actual stellar yields, star formation and gas-infall history of
individual galaxies, and on the Initial Mass Function (IMF) (Prochaska
et al$.$ 2002; Pilyugin, Thuan,
\& V\'{\i}lchez 2003; K\"oppen \& Hensler 2005; Moll\'{a} et al$.$ 
2006). In the case of the XUV disks these properties may be different
from those found in the HII regions of actively star-forming dwarfs
and optical spiral disks, places where metal abundances are
traditionally measured. For example, according to some of the
scenarios proposed to explain the lack of line emission in XUV disks
(Meurer et al$.$ 2004), the IMF in these regions might have a lack of
very massive stars. Such a top-light IMF would result in both the
oxygen and nitrogen to be released by intermediate-mass stars which
would led to a high N/O abundance ratio, higher than that adopted by
Kewley \& Dopita (2002) at low oxygen abundances. Should the actual
N/O abundance ratio in XUV disks behave differently from what these
authors assume some of the conclusions drawn above from the analysis
of secondary indictors could change. A more extensive discussion on
the possible N/O abundance ratios in these XUV disks is given in
Section~\ref{cloudy.no}.

Moreover, some of the conclusions on the analysis of the secondary
oxygen-abundance indicators are valid only if the ionization parameter
in these regions is in the range considered by standard
photoionization models such as those of Kewley \& Dopita (2002). Here
we should remind the reader that it has not been proved yet that
photoionization models of typical HII regions adequately reproduce the
physical conditions present in these, newly-discovered, extended UV
disks.

This paper is a first attempt to study the physical properties of the
ionized gas in these regions, so we face many limitations that are
inherent to previous empirical and theoretical studies mainly focused
on the analysis of HII regions in the main body of spiral and
irregular galaxies. In particular, the ionization parameter in the HII
regions of the XUV disks could be either significantly higher or lower
than in Galactic HII regions (due, for example, to a hotter/cooler
mean stellar population). They could also show a different density
structure than that found in typical HII regions and perhaps similar
to that of the disks of Low Surface Brightness (LSB) galaxies, where
the lower ISM pressure is thought to result in a more simple density
structure (Mihos, Spaans, \& McGaugh 1999). Moreover, single-star
photoionization models might be more appropriate when trying to
reproduce the conditions and output spectra of these regions (see
Section~\ref{cloudy}). And finally, the N/O abundance ratio in XUV
disks may be different from that measured elsewhere (see
Section~\ref{cloudy.no}).

\subsubsection{NGC~4625}

For NGC~4625 the [NII]/[OII] ratio in the emission-line regions of its
XUV disk are all below or very close to $-$1.0 (in logarithmic scale),
with the highest value being $-$0.94. Therefore, according to the
photoionization models of Kewley
\& Dopita (2002), the metallicities are necessarily below solar 
(for any ionization parameter between 5$\times$10$^{6}$ -or lower- and
3$\times$10$^{8}$\,cm\,s$^{-1}$). We remind the reader that we are
using 12+log(O/H)$_{\odot}$=8.69 (Allende Prieto et al$.$
2001). Moreover, the $B$-band luminosity-metallicity relationship
recently derived by Salzer et al$.$ (2005) indicates that for the
luminosity of NGC~4625 (M$_{B}$$\simeq$$-$17.4\,mag) the vast majority
of the local star-forming galaxies show oxygen abundances clearly
below 12+log(O/H)$<$8.5.

Therefore, it is reasonable to conclude that the low-metallicity
branch should be adopted for the XUV disk of NGC~4625, which is bluer
and probably less chemically evolved than the overall galaxy stellar
population (Gil de Paz et al$.$ 2005). It is worth emphasizing here
that the [NII]/[OII] ratio depends only weakly on the ionization
parameter. This conclusion would still be valid even if the N/O
abundance ratio in these regions is higher than log(N/O)=$-$1.5, the
constant value adopted by Kewley \& Dopita (2002) for primary-nitrogen
production.

In one of the regions of the XUV disk of NGC~4625 (NGC~4625: XUV 09),
the R23 value derived is somewhat larger than the maximum value of
this parameter for the corresponding [OII]/[OIII] ratio (or excitation
parameter, $P$), which results in an abnormally high value for the
oxygen abundance, in spite its low [NII]/[OII] ratio ($-$1.2 in the
log). This suggests that in this region the metal abundance is
probably close to the value for which the R23 ratio is maximuum
according to both photoionization models and empirical recipes
\{12+log(O/H)$\simeq$8.4 for
log([OII]/[OIII])$\simeq$0; see e.g$.$ Kobulnicky et al$.$ 1999\} and
that observational errors might have moved its value above this
maximuum. Therefore, in the particular case of NGC~4625: XUV 09 we
adopt 12+log(O/H)=8.4 as our best metallicity estimate. The PT05
calibration is of limited use in the case of NGC~4625, since all
regions analyzed in its XUV disk (except for NGC~4625: XUV 01) show
excitation parameters, $P$, well below 0.55, the minimum value given
by PT05 as giving reliable results. Its PT05-based oxygen abundances
are, consequently, extrapolated values.

\subsection{Photoionization modeling}
\label{cloudy}

\subsubsection{Physical parameters of the models}
\label{cloudy.parameters}

In order to study the physical conditions in the XUV disks we have
compared the observed line ratios with the predictions of a complete
set of $\sim$13,000 photoionization models covering a wide range of
physical properties both for the gas and for the source of ionizing
radiation. Calculations were performed with version 06.02 of CLOUDY,
last described by Ferland et al$.$ (1998).

First, we ran two sets of 3,262 simulations each using single massive
stars from the CoStar library of Schaerer \& de Koter (1997) with ZAMS
masses 20, 25, 40, 60, 85, and 120\,M$_{\odot}$, ages between 1 and
7\,Myr (or the lifetime of the star, whichever happens first) with a
step of 1\,Myr for stars $>$25\,M$_{\odot}$ and of 2\,Myr below that
mass, and two different stellar metallicities (Z$_{\odot}$ and
Z$_{\odot}$/5). The luminosity of the star was determined using the
relations given by Tout et al$.$ (1996). For the gas we adopted a
spherical distribution with constant density 1, 10, 10$^{2}$, and
10$^{3}$\,cm$^{-3}$ and a inner radius between 10$^{16}$\,cm to
10$^{17.5}$\,cm (in steps of 0.5\,dex). The outer radius was set to
10$^{21}$\,cm ($\sim$325\,pc) in order to ensure that all modeled HII
regions were ionization bounded. The metal abundance of both gas and
dust grains was allowed to simultaneously vary between Z$_{\odot}$/40
and 2$\times$Z$_{\odot}$, with values (1/40, 1/20, 1/10, 1/5, 3/10,
2/5, 1/2, 3/5, 4/5, 1, 1.5, and 2$\times$Z$_{\odot}$). The first set
of realizations was run adopting a constant solar N/O abundance ratio
\{log(N/O)$_{\odot}$=$-$0.76\} while for the second one we adopted a
constant value of log(N/O)=$-$1.5 for oxygen abundances below
12+log(O/H)=8.45 and a N/O rising linearly with O/H above that value,
i.e$.$ a behavior similar to that adopted in the photoionization
models developed by Kewley \& Dopita (2002). We stopped the
calculation when the gas temperature reached 3,000\,K, value at which
the contribution of the gas to the optical line emission is
negligible.

As we commented in Section~\label{results.lineratios} the weak line
fluxes of these HII regions indicate that they are most likely powered
by single, at a maximum, a few massive stars and not by a massive
stellar cluster. In order to further verify this result we will also
ran two sets of models (with two different N/O abundance ratios) of
3,456 CLOUDY realizations each where the gas was ionized by a stellar
cluster with a mass of 10$^{3}$\,M$_{\odot}$ (see Gil de Paz et al$.$
2005). For the stellar continuum we used the predictions of the
Starburst99 population systhesis models (Leitherer et al$.$ 1999) for
metallicities between Z$_{\odot}$/20 and 2$\times$Z$_{\odot}$ (with
values 1/20, 1/5, 2/5, 1, 2$\times$Z$_{\odot}$; i.e$.$ those of the
Geneva tracks used in these realizations) and ages in the range
1-9\,Myr (with steps of 0.1\,Myr), assuming the IMF of Kroupa, Tout,
\& Gilmore (1993). We made use of the Starburst99 default (recommended) 
stellar atmospheres of Pauldrach/Hillier (Pauldrach 2005, Hillier
2003, Lejeune, Cuisinier, \& Buser 1997). The spatial distribution and
chemical composition of the gas and dust grains were those used for
the single-star photoionization realizations described above. We used
a somewhat finer grid of gas densities (0.5\,dex steps instead of
1\,dex) in order to better explore the ionization parameter. Thanks to
the homology relations of photoionization models with the ionization
parameter the use of a fixed stellar mass for the cluster should not
strongly limit the range of possible line ratios derived as long as
the density of the gas is adequately sampled.

For $\sim$10\% of the starburst models the outer radius
(10$^{21}$\,cm, i.e$.$ 325\,pc) was reached before the temperature was
below 3,000\,K. This happens only for very low densities
(10$^{0.0-0.5}$\,cm$^{-3}$) and only for low gas metallicities or for
high metallicities when the ionizing starburst is young
($\leq$3\,Myr). None of the best-fitting solutions derived correspond
to one of these ``density-bounded'' models.

The best-fitting set of models in each case was determined by
$\chi^2$-minimization of the extinction-free line intensities
normalized to H$\beta$ in logarithmic scale. We adopted an average
error of 0.10\,dex in all line ratios, except for
[OII]$\lambda\lambda$3726,3729\AA\AA/H$\beta$ and
[SII]$\lambda\lambda$6717,6731\AA\AA/H$\beta$ where errors of 0.15 and
0.20\,dex were assumed, respectively (see
Section~\ref{calibration}). We checked that the intrinsic
H$\alpha$/H$\beta$ ratio predicted by these realizations was similar
(within $\sim$5\%) to the case-B recombination coefficient assumed as
part of our procedure for correcting the observed line ratios for
extinction. Only the properties of regions for which the
[OII]$\lambda\lambda$3726,3729\AA\AA/H$\beta$,
[OIII]$\lambda$5007\AA/H$\beta$, and [NII]$\lambda$6584\AA/H$\beta$
line ratios are available were determined. When the
[SII]$\lambda\lambda$6717,6731\AA\AA/H$\beta$ line ratio was
available, we included it also in the minimization. We note that the
S/O abundance ratio is approximately constant with the oxygen
abundance and similar to the solar value (see Kehrig et al$.$ 2006 and
references therein).

\subsubsection{N/O abundance ratio}
\label{cloudy.no}

The N/O abundance ratio is a function of the stellar yields, the star
formation and gas-infall histories of galaxies, and of the IMF. Most
galaxies follow a relatively well defined sequence in the N/O-O/H
diagram with a constant value of the N/O ratio for oxygen abundances
below 12+log(O/H)=8.5, which suggests a primary origen for the
nitrogen, and a N/O increasing linearly with O/H above that value,
which is explained in terms of a secondary production of nitrogen (van
Zee et al$.$ 1998a; Liang et al$.$ 2006; see also Moll\'{a} et al$.$
2006 for a theoretical review on the subject). However, as we
commented in Section~\ref{strong.m83}, both the history and IMF of the
outermost regions of spiral disks (such as those analyzed here) may be
quite different from those of the galaxies where nitrogen and oxygen
abundances have been traditionally measured. Due to this limitation we
decided to consider two possible N/O abundance ratios in our
photoionization models: one similar to that used by Kewley \& Dopita
(2002) and that is based on the results of van Zee et al$.$ (1998a)
and other where N/O is constant throughout the entire range in oxygen
abundance and equal to the solar value. The main difference between
the predictions of these two sets of models is the intensity of the
[NII]$\lambda$6584\AA\ line. This is due to the fact that this line or
other nitrogen lines have only a small contribution to the cooling of
the gas, so a change in the N/O abundance ratio does not alter
significantly the temperature structure of the nebula.

Tables~4 through 7 show that in the case of M\,83 the models where the
N/O is assumed to be constant and equal to the solar value provide a
better agreement to the data than those obtained using a O/H-scaled
N/O abundance ratio. The observed increase in the best-fitting reduced
$\chi^2$ for M\,83 is consequence of the [NII]$\lambda$6584\AA\ line
in the models being too weak for any set of physical parameters of the
nebula compatible with the observed oxygen line ratios. This can be
clearly seen in Table~8 where we compare the observed line ratios (in
logarithmic scale) with those predicted by the best-fitting
photoionization model obtained for each combination of ionization
source and N/O ratio. The models with O/H-scaled N/O abundance ratio
systematically underpredict the value of the extinction-corrected
[NII]$\lambda$6584\AA/H$\beta$ line ratio.

This result suggests that the actual N/O abundance ratio in the XUV
disk of M\,83 might be higher than that obtained in actively
star-forming dwarf galaxies and in the optical disks of spiral
galaxies at low oxygen abundances
\{log(N/O)=$-$1.5\}. There are several possible explanations for such
a high N/O, e.g$.$ the presence of a dominant evolved stellar
population (Pilyugin et al$.$ 2003; Moll\'{a} et al$.$ 2006) or a
top-light IMF. The former scenario seems unlikely considering the
relatively blue colors measured in the outer parts of these and other
spiral disks (Mu\~noz-Mateos et al$.$ 2006; see also
Section~\ref{SFH}). Alternatively, an episodic star formation history
with the last episode of star formation taking place several Gyr ago
(i.e$.$ time long enough for the release of nitrogen by 
intermediate-mass stars) could also result in relatively high N/O ratios.

Another possible scenario involves the infall of pristine gas in
regions with moderate (close to solar) oxygen abundance where the
nitrogen production is already secondary. This gas infall would drive
the system into a region of lower oxygen abundance without changing
the N/O ratio (K\"oppen \& Hensler 2005). This scenario has been also
proposed to explain the large dispersion in the N/O ratio at oxygen
abundances 12+log(O/H)$\geq$7.9 (including some objects with high
values of N/O) recently found in a sample of dwarf irregular galaxies
with low ionization parameters (van Zee \& Haynes 2006).

In the case of NGC~4625 the results obtained for the two different N/O
ratios considered are comparable except for region NGC~4625: XUV 05
where the solar N/O yields a better fit and for regions NGC~4625: XUV
09 and 12 where a O/H-scaled N/O ratio is favored. However, for most
regions (all except NGC~4625: XUV 01) the line fluxes predicted by the
best-fitting single-star models with a O/H-scaled N/O ratio are
systematically lower than the lower limits measured by up to a factor
of 40 (see Tables~2 \& 5).

For the remaining of the paper we will focus on the results obtained
for the set of models with solar N/O abundance ratio (Tables~4 \& 6)
since they provide a better fit to the data overall. 

\subsubsection{Best-fitting single-star models}
\label{cloudy.ss}

In Table~4 we show the parameters of the best-fitting single-star
photoionization model with solar N/O ratio. We find that the lowest
reduced $\chi^2$ achieved are not very different from 1 for most of
the regions, which indicates that these single-star photoionization
models provide a good fit to the line ratios measured. The fact that
they are not much smaller than 1 also indicates that we are not
over-fitting our data. The high reduced $\chi^2$ found in some of the
regions might be due to these regions being powered by a few stars of
different effective temperature or by a larger composite stellar
population such a stellar cluster (see below). We have estimated the
uncertainties in the properties derived by determining the minimum
(subscripts in Tables~4 \& 5) and maximum (superscripts in Tables~4 \&
5) values of the each parameter that result in values of the absolute
(i.e$.$ not reduced) $\chi^2$ that difer from the minimum $\chi^2$ by
less than $\Delta\chi^2$=1 (see Avni 1976). This yields the 68.3\%
confidence intervals in each parameter when they are considered
separately. Table~4 shows that while the ages and masses of the
ionizing stars and the metal abundance of the gas are relatively well
constrained, the inner radius of the gas distribution is very poorly
constrained by our measurements.

In roughly half of the regions (NGC~5236: XUV 01, 07, 10, 11, 17, and
NGC~4625: XUV 01, 05, 09) the H$\beta$ fluxes predicted by the
best-fitting model are comparable or a factor of a few above
(explicable by slit-aperture effects) the fluxes measured through the
slits (Table~2). There are some cases (NGC~5236: XUV 09, 12, 14, and
NGC~4625: XUV 02, 06, 07) where the measurements are above the
models but only by a factor $\sim$2-5$\times$. These HII regions could
well be powered by a very few massive stars. Finally, in a few regions
(NGC~5236: XUV 13, 20, 21, 22, and NGC~4625: XUV 06, 12) the slit
H$\beta$ fluxes are more than a factor of 10 above those predicted by
the models suggesting that the source of the ionizing radiation is a
relatively massive stellar cluster. This was expected in the case of
regions NGC~5236: XUV 20, 21, and 22 since they are very luminous
regions in the optical disk of M\,83.

In the case of M\,83 the ZAMS masses derived for the ionizing stars of
the XUV regions are in the range 20-40\,M$_{\odot}$ with a
best-fitting age of 1 to 3\,Myr. With respect to the properties of the
gas, the densities and inner radii show a wide range of values between
1-10$^{3}$\,cm$^{-3}$ and 10$^{16}$-10$^{17.5}$\,cm, respectively. The
best-fitting single-star models for NGC~4625 yield lower gas densities
and somewhat higher temperatures and masses for the ionizing stars
than those of M\,83.

The gas metal abundances nicely agree with those derived in
Section~\ref{abundances} using the R23 ratio. Indeed, while M\,83
shows a strong metallicity gradient with abundances between
Z$_{\odot}$/10 and Z$_{\odot}$/4 in the outer disk and close or even
above solar in the optical disk, the XUV-disk regions of NGC~4625
(except for NGC~4625: XUV 09, with $\chi^2$$>$5) show best-fitting
metallicities that are either Z$_{\odot}$/5 or Z$_{\odot}$/10.

In light of these results we can now confidently say that metal
abundances as low 12$+\log{\mathrm{(O/H)}}$=7.86 (regions NGC~5236:
XUV 01 and NGC~5236: XUV 10) and 12$+\log{\mathrm{(O/H)}}$=7.94
(NGC~4625: XUV 01) are found in the XUV disks of M\,83 and NGC~4625,
respectively. These metallicities are comparable to the lowest oxygen
abundances found by van Zee et al$.$ (1998b;
12$+\log{\mathrm{(O/H)}}$=7.92 in NGC~5457) and Ferguson, Gallagher,
\& Wyse (1998; 12$+\log{\mathrm{(O/H)}}$=7.95 in NGC~1058) in the outermost
regions of a sample of spiral galaxies. Note that in the case of
Ferguson et al$.$ (1998) only galaxies in Ferguson (1997) that showed
very extended HI and extreme outer-disk star formation were analyzed.

\subsubsection{Best-fitting starburst models}
\label{cloudy.sb}

Table~6 shows the properties and reduced $\chi^2$ of the best-fitting
starburst models with solar N/O ratio. The reduced $\chi^2$ values are
larger (i.e$.$ worse than) or, at best, comparable to those obtained
from the corresponding single-star photoionization model. This
indicates that for low stellar masses such as those of the XUV
complexes (Gil de Paz et al$.$ 2005) the shape of the ionizing
spectrum is different from that of a modeled composite stellar
population, which are usually built for massive star-forming regions
and starburst galaxies. In particular, even in those cases where the
presence of more than one ionizing star is suspected, stochastic
effects on the upper IMF might result in the ionizing radiation from
these regions to be dominated by only two or three of these stars,
which would lead to an ionizing spectrum different from that of a
low-mass stellar cluster with a continuous IMF, such as those
generated by the Starburst99 population synthesis
models. Unfortunately, the constraints imposed by the few line ratios
measured in this study are limited so the determination of the number
and masses of the ionizing stars present in each of these regions is
beyond our reach. High-resolution imaging observations using HST would
partially solve this problem by resolving individual massive stars at
the distance of the nearest XUV disks. As expected, younger
best-fitting starburst models typically correspond to regions where
the effective temperature of the best-fitting single-star model is
higher. The 68.3\% confidence intervals in metal abundance of some of
the regions in NGC~4625 are noticeably wider that those found using
single-star models. Again, we find that the inner radius of the dust
distribution is very poorly constrained.

We note that although for the low metal abundances of the regions
analyzed here the differences between the CoStar library and
Pauldrach/Hillier stellar atmospheres used in Starburst99 are not as
dramatic as for Z$\geq$Z$_{\odot}$ (Smith, Norris, \& Crowther 2002)
differences between the two models could also partially explain the
different best-fitting $\chi^2$ values achieved. Should that be the
case the massive stars present in the emission-line regions in the XUV
disk would likely have a harder ionizing spectrum (as in the CoStar
library; see Smith et al$.$ 2002) than the O-stars included in the
Starburst99 models.

However, we should emphasize here that the fact that single-star
models better reproduce the line ratios was suspected already based
uniquely on the low emission-line fluxes measured in these regions
(see Table~2). Unfortunately, as we commented in
Section~\ref{results.lineratios}, some of these spectroscopic fluxes
might be significantly affected by aperture effects, so our results
could not be based solely on that fact.

\subsection{Metallicity gradients}
\label{gradients}

The oxygen metal abundances of the HII regions in the disk of M\,83
are found to decrease with the galactocentric distance (Figure~7a). If
we assume that for regions NGC~5236: XUV 09 to NGC~5236: XUV 14 the
low-metallicity branch of the R23 calibration is the most appropiate
the best linear fit yields a metallicity gradient of
d[O/H]=$-$0.112\,dex\,kpc$^{-1}$ (d[O/H]=$-$0.069\,dex\,kpc$^{-1}$)
using the M91 (PT05) method, with
\begin{eqnarray}
12+\log{\mathrm{(O/H)}}=-0.112 \times r\mathrm{(kpc)}+9.36 ; \sigma=0.37\,\mathrm{dex}\ \ ({\rm M91})\\ 
12+\log{\mathrm{(O/H)}}=-0.069 \times r\mathrm{(kpc)}+8.60 ; \sigma=0.34\,\mathrm{dex}\ \ ({\rm PT05}) 
\end{eqnarray}
Here we have combined our metallicity measurements for the XUV disk
with those derived from the line ratios of HII regions in the optical
disk (Webster \& Smith 1983) using the same methodology as for the
XUV-disk abundances (open dots in Figure~7a). If the high-metallicity
branch of the R23 calibration is now assumed for regions NGC~5236: XUV
09 to NGC~5236: XUV 14 the metallicity gradient is much shallower,
\begin{eqnarray}
12+\log{\mathrm{(O/H)}}=-0.051 \times r\mathrm{(kpc)}+9.26 ; \sigma=0.16\,\mathrm{dex}\ \ ({\rm M91})\\ 
12+\log{\mathrm{(O/H)}}=-0.017 \times r\mathrm{(kpc)}+8.51 ; \sigma=0.25\,\mathrm{dex}\ \ ({\rm PT05}). 
\end{eqnarray}
Finally, if the branch of the R23 calibration is assigned based on the
metal abundance of the best-fitting CLOUDY single-star photoionization
model with solar N/O, i.e$.$ low-metallicity branch for all regions
except for NGC~5236: XUV 12, 14, 20, 21, and 22, the result is
\begin{eqnarray}
12+\log{\mathrm{(O/H)}}=-0.100 \times r\mathrm{(kpc)}+9.41 ; \sigma=0.30\,\mathrm{dex}\ \ ({\rm M91})\\ 
12+\log{\mathrm{(O/H)}}=-0.058 \times r\mathrm{(kpc)}+8.64 ; \sigma=0.30\,\mathrm{dex}\ \ ({\rm PT05}). 
\end{eqnarray}
These values are slightly stepper and shallower, respectively, than
the metallicity gradient measured for the Milky Way
(d[O/H]$_{\mathrm{MW}}$$\sim$$-$0.08\,dex\,kpc$^{-1}$; Boissier et
al$.$ 1999 and references therein) and than those derived for a sample
of spiral galaxies with oxygen abundances measurements at large radii
(van Zee et al$.$ 1998b, Ferguson et al$.$ 1998). If we now express the
metallicity gradient in units of the blue-band scale-length ($r_{d}$)
or D25 radius ($r_{25}$) we obtain $-$0.198\,dex/$r_{d}$ and
$-$1.01\,dex/$r_{25}$, respectively, in the case of the M91
calibration and $-$0.115\,dex/$r_{d}$ and $-$0.59\,dex/$r_{25}$ with
the PT05 calibration. These values are similar to the ``universal''
gradient of $-$0.2\,dex/$r_{d}$ (or $-$0.8\,dex/$r_{25}$) proposed by
Prantzos \& Boissier (2000, see also Garnett et al$.$ 1997) for
non-barred disk galaxies and within the range of values found by van
Zee et al$.$ (1998) and Ferguson et al$.$ (1998) in spirals with
outer-disk star formation.

In light of Figure~7a instead of a single linear fit, the radial
dependence of metal abundance seems to suggest a smooth metallicity
gradient below 10\,kpc and sharp decrease in metallicity from
$\sim$Z$_{\odot}$ to Z$_{\odot}$/5 (if the M91 calibration is used) at
that radius. This behavior could be explained if there exists a
cut-off radius beyond which the disk of M\,83 is significantly less
efficient in forming stars and it is allowed to evolve mostly
viscously (Clarke 1989). Note, however, that this might also be the
consequence of the large uncertainties expected for any R23-based
metallicity estimate at abundances close to half the solar value. A
larger number of metallicity measurements, specially in the range
10-15\,kpc, and more precise estimates of the metallicity using the
[OIII]$\lambda$4363\AA\ temperature-sensitive line would be required
to confirm (or rule out) this.

In the case of NGC~4625, where the low-metallicity branch can be
safely assumed, the best linear fits obtained are
\begin{eqnarray}
12+\log{\mathrm{(O/H)}}=-0.037 \times r\mathrm{(kpc)} + 8.44 ; \sigma=0.20\,\mathrm{dex}\ \ ({\rm M91})\\
12+\log{\mathrm{(O/H)}}=-0.017 \times r\mathrm{(kpc)} + 8.16 ; \sigma=0.39\,\mathrm{dex}\ \ ({\rm PT05}).
\end{eqnarray}
The metallicity gradient derived, $\Delta$[O/H]/$\Delta
R$=$-$0.037$\pm$0.032\,dex\,kpc$^{-1}$ (using the M91 calibration), is
very modest. Even after normalizing this value by the galaxy
scale-length or D25 radius, this is one of the shallowest gradients
measured in disk galaxies, $-$0.080\,dex/$r_{25}$. Note, however, that
NGC~4625 is significantly less luminous (M$_{B}$=$-$17.4\,mag) than
the spiral galaxies where metallicity gradients have been
traditionally measured.

The dust attenuation also seems to decrease with the galactocentric
distance in the case of M\,83 although with a very large dispersion
(Figure~7b). The fact that this scatter is present even when
neighboring regions are considered (e.g$.$ NGC~5236: XUV 07 and
NGC~5236: XUV 08) suggests a patchy dust distribution. In the case of
NGC~4625 (see Figure~7d) no obvious trend with the galactocentric
distance is found.

It is also remarkable that positive and relatively large values of
C(H$\beta$) are measured very far out into the disks, showing the
presence of dust in these galaxies at distances larger than 18\,kpc in
the case of M\,83 and more than 10\,kpc in NGC~4625 (see Popescu \&
Tuffs 2003 for an example of dust emission associated with an extended
HI disk).

\section{Discussion}
\label{discussion}

\subsection{Nature of the XUV disks: XUV complexes}

From the analysis carried out in this paper and from previous studies
on the extended-UV disk phenomenon (Thilker et al$.$ 2005a, Gil de Paz
et al$.$ 2005) is now clear that the XUV emission is due to young
stars associated with low-mass stellar associations located at large
galactocentric distances. The line ratios measured in those XUV
complexes showing line emission indicate the presence of UV ionizing
radiation being emitted locally in these regions. It is therefore
unlikely that the non-ionizing UV light responsible for the XUV
emission seen by GALEX is due to scattering by dust, like it is the
case of the polar regions of the starburst galaxy M~82 (Hoopes et
al$.$ 2005). Our measurements also exclude planetary nebulae (with
very high temperatures and highly-excited gas; see
Section~\ref{results.lineratios}) as being the agents responsible for
the observed XUV emission. Finally, a significant contribution by
blue-HB stars it is highly unlikely considering (1) the blue
NUV$-$optical colors of the XUV disks, (1) the high equivalent widths
measured in some XUV complexes and (2) the morphology (resembling
spiral arms) of the XUV emission; all unusual circumstances in objects
with UV emission dominated by evolved stars (globular clusters,
elliptical galaxies) but common in galaxies forming massive stars in
recent epoch.

Our results also support the conclusions of Gil de Paz et al$.$ (2005)
that suggested that the majority of the admittedly scarce
emission-line sources found in the XUV disks of these galaxies seem to
be powered by single stars, with masses between 20-40\,M$_{\odot}$ in
most cases. As a consequence of this, a stochastic treatment of the
IMF (e.g$.$ Cervi\~no 1998) is necessary in order to properly analyze
the properties of the stellar populations in these XUV disks and
perhaps in the outer edges of all spiral galaxies. In particular,
knowledge on the fraction of low-mass stellar complexes that, once
stochastic effects are properly accounted for, would form at least one
ionizing star is key to understanding the observed differences between
the UV and H$\alpha$ light profiles of XUV disks (see e.g$.$ Meurer et
al$.$ 2004).

In this sense, it has been recently argued (Boissier et al$.$ 2006)
that the lack of H$\alpha$ emission in the azimuthally-averaged
profiles of spiral galaxies beyond a given radius, which is commonly
associated with the existence of a intrinsic star formation threshold
(e.g$.$ Martin \& Kennicutt 2001), could instead be a consequence of
the fact that at those low levels of star formation the number of
ionizing stars expected to be found (at a given time) inside one of
these annuli of the radial profile is less than unity. Under these
circumstances, the chances of catching one of these ionizing stars are
very low, leading to an effective lack of HII regions in most cases,
but not necessarily to a lack of {\it time-integrated} star formation.

\subsection{Nature of the XUV disks: Extended HI, cause or consequence?}

Another property of the XUV disks is that they are always found in
galaxies with very extended HI distributions (see Tilanus
\& Allen 1993 in the case of M\,83 and Bush \& Wilcots 2004 for
NGC~4625). This is not surprising since gas is of course a necessary
ingredient for the formation of the stars responsible of the UV
emission and the neutral-to-molecular gas ratio is thought to increase
with the galactocentric radius in field spiral galaxies (e.g. Kohta,
Naomasa, \& Kuno 2001). However, the fact that significant amounts of
HI (compared with their stellar content) are still found at the
positions of these UV complexes suggests that the efficiency in
converting neutral gas into stars is relatively low and that, under
the right circumstances, these XUV disks could maintain the observed
level of star formation for a cosmologically significant period of
time. The SFR per unit area of these XUV disks can be obtained from
the azimuthally-averaged FUV-surface-brightness after a correction for
extinction of A$_{\mathrm{FUV}}$=1\,mag is adopted\footnote{This
corresponds to an average observed (FUV$-$NUV) color of 0.3\,mag and
the relationship between (FUV$-$NUV) color and A$_{\mathrm{FUV}}$
given by Boissier et al$.$ (2006). Note that UV magnitudes are in AB
scale and optical magnitudes are referred to Vega throughout the
text.}. If we now use the gas-surface-density profile of Thilker et
al$.$ (2005a) the consumption timescale derived for M\,83 is
$\geq$6\,Gyr at all radii, only considering the HI. In the case of
NGC~4625 the consumption timescale is $\sim$3\,Gyr in the innermost
parts of the XUV disk and rises up to values of the order of a Hubble
time in the outer regions (Gil de Paz et al$.$ 2005).

Alternatively, it could be argued that the presence of an extended HI
disk is a consequence of the XUV emission instead of a necessary
pre-existing condition (Allen 2002). Under this scenario, the extended
HI emission would be the result of the dissociation of H$_{2}$ in
photodissociation regions (PDRs) by the UV photons emitted by the same
newly-formed stars responsible for the XUV emission. 

However, in most of these XUV disks the complexes responsible for the
FUV emission cover only a fraction of the area of the HI
disk. Therefore, in order for this scenario to be valid in the case of
the XUV disks the time for the reformation of the H$_2$ molecule on
the surface of dust grains from atomic hydrogen, $\tau_{form}$, should
be larger than, or at least comparable to, the average time a
particular region in the disk is embedded in FUV radiation. This
condition can be written as
\begin{equation}
\tau_{form} \gtrsim {{A_{\mathrm{HI}}}\over{A_{\mathrm{FUV}}}} \times \tau_{\mathrm{FUV}}
\label{eq7}
\end{equation}
where $A_{\mathrm{HI}}$ is the area of the XUV disk where HI emission
is detected, $A_{\mathrm{FUV}}$ is the area of the disk reached by the
FUV photons (responsible for the H$_2$ photodissociation in PDRs), and
$\tau_\mathrm{FUV}$ is the timescale of the FUV emission, which is
the order of a few hundred Myr for an instantaneous burst.

The H$_2$ reformation rate can be obtained using the expression
given by Hollenbach \& McKee (1979),
\begin{equation}
R_{\mathrm{H}_2} \equiv {{\mathrm{d}n_{\mathrm{H}_2}}\over{\mathrm{d}t}} = {{1}\over{2}} n_{\mathrm{H}} v_{\mathrm{H}} \sigma \xi \eta n_{\mathrm{g}}
\end{equation}
where $n_{\mathrm{H}}$ is the number density of hydrogen atoms (in
atomic form), $v_{\mathrm{H}}$ is their themal speed, $\sigma$ is the
cross section of the grain, $\xi$ is the sticking coefficient, $\eta$
is the probability of bond formation once two hydrogen atoms encounter
each other and $n_{\mathrm{g}}$ is the number density of dust grains
in the ISM. Here we assume that the contribution of the H$^{-}$
process to the formation of H$_2$ is negligible. From the expression
of the reformation rate given above, the reformation time, defined as
the time necessary for half of the total hydrogen atoms to be in
molecular form, can be written as
\begin{equation}
\tau_{form} = {{\ln 2}\over{v_{\mathrm{H}} \sigma \xi \eta n_{\mathrm{g}}}}
\end{equation}
For simplicity, we assume a gas temperature of 70\,K (Hughes,
Thompson, \& Colvin 1971), which yields a thermal speed for the
hydrogen atoms of 1.2$\times$10$^{5}$\,cm\,s$^{-1}$ (Spitzer
1978). The cross section of the grain is computed as $\pi a^2$, where
$a$ is the typical radius of the grains in the ISM (0.1\,$\mu$m; Jones
\& Merrill 1976). The sticking coefficient is adopted to be $\xi$=0.3
and $\eta$=1 (Vidali et al$.$ 2005; Hollenbach \& Salpeter 1971). The
number density of grains is estimated to be
2\,10$^{-12}$$\times$$n_{\mathrm{H}}$ after an average grain material
density of 2\,g\,cm$^{-3}$ (Draine
\& Lee 1984) and the Galactic ISM gas-to-dust mass ratio are adopted ($R_{g/d}$$\sim$100;
Knapp \& Kerr 1974). This yields a reformation time of
\begin{equation}
\tau_{form} \mathrm{(Myr)} \approx {{10^3}\over{n_{\mathrm{H}}}}
\label{eq10}
\end{equation}
where $n_{\mathrm{H}}$ is expressed in cm$^{-3}$. Allen (2002) gives a
similar expression although his reformation timescales are somewhat
smaller than those given by Equation~\ref{eq10}. This is in part due
to the fact that he uses parameters typical of PDRs while in our case
the reformation of the H$_2$ is expected to take place in a more
rarefied ISM.

For XUV disks where the FUV emission covers a high fraction of the
disk (like in the case of the inner XUV disk of NGC~4625; Gil de Paz
et al$.$ 2005) the reformation time might fulfill the condition
imposed by Equation~\ref{eq7} as long as the gas is not very dense. On
the other hand, in M\,83, where the FUV emission covers only
$\sim$10\% of the extended HI disk, the gas density required for the
hydrogen to stay in atomic form is very low
($n_{\mathrm{H}}$$<$1\,cm$^{-3}$). We note that the Galactic
gas-to-dust ratio assumed above corresponds to a ISM with a higher
abundance of metals and probably dust grains than that expected to be
present in the XUV disk of M\,83, where a larger $R_{g/d}$ might exist
(see Galliano et al$.$ 2003, 2005). Moreover, the size of the PDRs
might be somewhat larger than that of the FUV-bright regions seen in
the GALEX images, which would lead to a smaller value of
$A_{\mathrm{HI}}/A_{\mathrm{FUV}}$.

Therefore, we cannot definitively exclude the possibility that a
significant fraction of the extended HI associated with the XUV disks
is a consequence of the photodissociation of H$_2$ in PDRs. 

A more detailed analysis of the relation between gas and recent star
formation in XUV disks, including a study of the star formation
threshold and the star formation law, will be the subject of a future
paper.

\subsection{Past and future evolution of the XUV disks}
\label{SFH}

The very low surface brightnesses of these XUV disks at optical
wavelengths and, consequently, the very blue UV-optical colors of
their stellar populations led to the suggestion that these might be
the first generations of stars being formed in these regions (Madore
et al$.$ 2004, Gil de Paz et al$.$ 2005). However, the oxygen
abundances derived for the emission-line regions of M\,83 and NGC~4625
(Z$\sim$Z$_{\odot}$/5-Z$_{\odot}$/10) indicate that in spite of being
relatively unevolved chemically these are not the first stars to have
formed in the outermost regions of these galaxies. Some star formation
activity must have taken place in these regions in the past in order
to enrich the ISM to its present levels.

There are many ways to produce such an enrichment of the ISM: 

\begin{itemize}
\item Low-level continuous star formation: If the star formation history 
of the XUV disks has been similar to that of young spiral disks during
their early formation at high redshift this level of chemical
enrichment could have been reached in a more or less continuous way
only $\sim$1\,Gyr after the first stars formed (see e.g$.$ Boissier \&
Prantzos 2000). A scenario where the first stars formed much earlier
than one billion years ago and have maintained the same level of star
formation is unlikely since that would result in redder colors than
those measured in the XUV disks. In particular, if we correct the
color profiles of NGC~4625 given in Gil de Paz et al$.$ (2005) for
extinction using the relation between the (FUV$-$NUV) color and
A$_{\mathrm{FUV}}$ found by Boissier et al$.$ (2006) for the disks of
spiral galaxies (assuming the Galactic extinction law) the corrected
(NUV$-B$) color would range between 0.4\,mag (inner XUV disk) and
$-$0.5\,mag (outer XUV disk). For this range of colors any stellar
population with an approximately constant star formation rate (and
Z=Z$_{\odot}$/5) would necessarily be younger than 1\,Gyr (e.g$.$
Bruzual \& Charlot 2003). On the other hand, the corrected ($B-R$)
color is approximately flat with a value of $\sim$0.6-0.7\,mag. This
value is consistent with a 2-5\,Gyr-old continuously-forming stellar
population or with a 500\,Myr-old instantaneous burst. However, since
the photometry errors in ($B-R$) are large compared with its age
sensitivity, significantly younger (or older) ages are also possible.

As noted above, the amount of gas in the XUV disks of these systems is
large enough to maintain the current level of star formation for
several Gyr and up to a Hubble time depending on the XUV disk and the
region of the XUV disks considered, especially once the contribution
of a possible molecular component to the total gas mass is
included. It is therefore conceivable that these galaxies can keep
forming stars in their outermost regions at the current rate for a few
Gyr and that the optical disk will steadily grow thanks to the
accumulation of the light from the resulting long-lived low- and
intermediate-mass stars.

\item Episodic star formation: Alternatively, an episodic star formation 
with brief epochs of active star formation followed by long quiescent
periods could also have led to this level of chemical enrichment. If
this were the case, other (now quiescent) galaxies might have gone
through one or several of these episodes during their lifes. Thilker
et al$.$ (2005b) have recently proposed that the so-called {\it
anti-truncated} disks identified by Erwin, Beckman, \& Pohlen (2005)
in early-type spirals could be the evolved counterparts of the massive
XUV-disk population. The acquisition of deep ground-based
multi-wavelength surface photometry and single-star color-magnitude
diagrams would greatly benefit the determination of the star formation
history of XUV disks along with its possible connection with the
anti-truncated disks. Deeper optical spectroscopy of some of these
regions with the aim of determining more precise O/H and N/O
measurements should also provide additional clues on the star
formation histories of XUV disks. Some of these efforts are currently
underway.
\end{itemize}

\section{Conclusions}
\label{conclusions}

Analysis of the emission-line spectra of a sample of
H$\alpha$-selected regions in the extended-UV disks of M\,83 and
NGC~4625 allow us to draw the following conclusions regarding the
properties of these disks:

\begin{itemize}

\item The line ratios measured in the emission-line regions of the XUV 
disks of M\,83 and NGC~4625 show that their ionizing radiation is
emitted by young stars and neither due to scattering by dust nor to
hot evolved stars, such as blue-HB or post-AGB stars.

\item The metal abundances derived, both from the R23 parameter and by comparison of multiple line ratios with the predictions of photoionization
models, are non primordial with oxygen abundances between
Z$_{\odot}$/10 and Z$_{\odot}$/4 in NGC~4625. The oxygen abundance in
M\,83 increases from $\sim$Z$_{\odot}$/10 in the outermost regions of
the XUV disk to close-to-solar (or even super-solar if the M91
calibration is adopted) abundances in the optical disk. The same
behavior is observed for the radial distribution of the dust
extinction: it is approximately flat in the case of NGC~4625 and shows
a progressive decrease with increasing galactocentric distance
(although with a large scatter) in M\,83. The comparison of the
observed line ratios with the model predictions favors a high N/O
abundance ratio for the XUV disk of M\,83. This could be due to the
infall of pristine gas in regions of secondary nitrogen production or
be the natural outcome of a top-light IMF.

\item The emission-line luminosities and ratios measured indicate that 
most of the HII regions analyzed are photoionized by single massive
stars and only a few are powered by a stellar cluster. This fact,
along with the low azimuthally-averaged star formation rates derived,
suggests that stochastic effects both on the upper IMF (Cervi\~no
1998) and on the notion of star formation threshold (Boissier et al$.$
2006) should not be ignored when studying the properties of the outer
edges of spiral galaxies in general and of XUV disks in particular.

\item The amount of gas in these XUV disks is enough to maintain the 
current level of star formation for several Gyr and, in some cases, up
to a Hubble time. We cannot exclude the possibility that the HI
associated with these disks may be the product of the dissociation of
H$_2$ in PDRs by the FUV radiation emitted by the stars formed during
the current and recent episodes of XUV emission.

\item The observed colors and the level of chemical enrichment measured 
suggest that these XUV disks have experienced either (1) a continuous
star formation for the last billion years or less, or (2) a episodic
star formation history with XUV episodes followed by quiescent
periods. Should the latter scenario be correct, other (perhaps all)
spiral galaxies might have gone through one or several of these
episodes.

\end{itemize}

\acknowledgments

GALEX (Galaxy Evolution Explorer) is a NASA Small Explorer, launched
in April 2003. We gratefully acknowledge NASA's support for
construction, operation, and science analysis for the GALEX mission,
developed in cooperation with the Centre National d'Etudes Spatiales
of France and the Korean Ministry of Science and Technology. We thank
the anonymous referee for his/her constructive comments that have
considerably improved the content of the paper. AGdP is financed by
the MAGPOP EU Marie Curie Research Training Network and partially by
the Spanish Programa Nacional de Astronom\'{\i}a y Astrof\'{\i}sica
under grant AYA2003-01676. We thank Judith Cohen for kindly providing
her H$\alpha$ filter for COSMIC. We are also thankful to Sergio
Gonz\'alez and Wojtek Krzeminski for carrying out the imaging
observations at Las Campanas 40-inch telescope.

{\it Facilities:} \facility{GALEX, Magellan:Baade (IMACS), Hale (COSMIC), Swope (Direct CCD)}

\clearpage
\begin{deluxetable}{lcccccccc}
\rotate 
\tabletypesize{\footnotesize}
\tablecaption{Summary of the imaging observations}
\tablecolumns{8}
\tablewidth{0pt}
\tablehead{
\colhead{Band} & \colhead{Date} & \colhead{Exposure} & \colhead{Instrument} & \colhead{Detector} & \colhead{Telescope} & \colhead{Scale} & \colhead{Seeing} \\
\colhead{}       & \colhead{} & \colhead{(seconds)}  &  \colhead{} & \colhead{} & \colhead{} & \colhead{(''/pixel)} & \colhead{(arcsec)} \\
\colhead{(1)} & \colhead{(2)} & \colhead{(3)} & \colhead{(4)} & \colhead{(5)} &\colhead{(6)} & \colhead{(7)} & \colhead{(8)}
}
\startdata
\cutinhead{M\,83}
$U$       & 2005/03/10-14 & 10800 & Direct CCD & 2048$\times$3150 SITe~3 CCD & Swope 1-m, LCO & 0.434 & 1.5\\
$R$       & 2005/03/10-14 &  2700 & Direct CCD & 2048$\times$3150 SITe~3 CCD & Swope 1-m, LCO & 0.434 & 1.1\\
H$\alpha$ & 2005/03/14    & 10200 & Direct CCD & 2048$\times$3150 SITe~3 CCD & Swope 1-m, LCO & 0.434 & 1.0\\
\cutinhead{NGC~4625}
$B$       & 1995/05/28 \& 1995/12/24 & 2400 & PFCU   & 1124$\times$1124 TEK~3 CCD & INT 2.5-m, La Palma & 0.59 & 1.4\\
$R$       & 1995/05/28 \& 1995/12/27 & 2200 & PFCU   & 1124$\times$1124 TEK~3 CCD & INT 2.5-m, La Palma & 0.59 & 1.3\\
H$\alpha$ & 2004/08/20               &  800 & COSMIC & 2048$\times$2048 SITe CCD  & Hale 5-m, Palomar & 0.40 & 1.0 \\
\enddata
\end{deluxetable}

\begin{deluxetable}{lccccr}
\tablecaption{XUV-regions parameters}
\tablecolumns{5}
\tablewidth{0pt}
\tablehead{
\colhead{XUV region name} & \colhead{RA(2000)}   & \colhead{DEC(2000)}   & \colhead{f$_{\mathrm{H}\beta}$}& \colhead{f$_{\mathrm{H}\beta,0}$} & \colhead{$d$}\\
\colhead{}            &  \colhead{(h:m:s)}    & \colhead{(d:m:s)}      &  \colhead{(erg\,s$^{-1}$\,cm$^{-2}$)} & \colhead{(erg\,s$^{-1}$\,cm$^{-2}$)} & \colhead{(kpc)}\\
\colhead{(1)} & \colhead{(2)} & \colhead{(3)} & \colhead{(4)} & \colhead{(5)} & \colhead{(6)}
}
\startdata
\cutinhead{M\,83 (NGC~5236)}
NGC~5236: XUV 01 & 13:36:55.149 & $-$30:05:47.51 & 1.11$\times$10$^{-15}$ & 1.11$\times$10$^{-15}$ & 18.2\\
NGC~5236: XUV 02 & 13:36:58.720 & $-$30:05:59.52 & 1.11$\times$10$^{-16}$ & 2.39$\times$10$^{-16}$ & 18.4\\
NGC~5236: XUV 03 & 13:37:00.371 & $-$30:05:54.83 & 7.91$\times$10$^{-17}$ & 2.07$\times$10$^{-16}$ & 18.3\\
NGC~5236: XUV 04 & 13:36:56.222 & $-$30:06:53.40 & 1.60$\times$10$^{-16}$ & 1.60$\times$10$^{-16}$ & 19.6\\
NGC~5236: XUV 05 & 13:36:59.284 & $-$30:01:01.85 & 5.78$\times$10$^{-17}$ & 3.33$\times$10$^{-16}$ & 11.9\\
NGC~5236: XUV 06 & 13:37:01.676 & $-$30:00:54.14 & 5.52$\times$10$^{-17}$ & 5.52$\times$10$^{-17}$ & 11.7\\
NGC~5236: XUV 07 & 13:37:05.182 & $-$30:00:14.37 & 7.06$\times$10$^{-17}$ & 3.02$\times$10$^{-16}$ & 10.9\\
NGC~5236: XUV 08 & 13:37:06.470 & $-$29:59:56.73 & 3.97$\times$10$^{-17}$ & 5.02$\times$10$^{-17}$ & 10.6\\
NGC~5236: XUV 09 & 13:37:05.001 & $-$29:59:46.05 & 1.07$\times$10$^{-15}$ & 1.57$\times$10$^{-15}$ & 10.3\\
NGC~5236: XUV 10 & 13:37:05.242 & $-$29:59:33.30 & 3.28$\times$10$^{-16}$ & 4.57$\times$10$^{-16}$ & 10.0\\
NGC~5236: XUV 11 & 13:37:08.196 & $-$29:59:19.67 & 4.05$\times$10$^{-16}$ & 4.05$\times$10$^{-16}$ & 9.9\\
NGC~5236: XUV 12 & 13:36:58.772 & $-$29:59:20.82 & 1.85$\times$10$^{-16}$ & 9.46$\times$10$^{-16}$ & 9.7\\
NGC~5236: XUV 13 & 13:36:58.523 & $-$29:59:24.54 & 3.70$\times$10$^{-15}$ & 6.93$\times$10$^{-15}$ & 9.8\\
NGC~5236: XUV 14 & 13:37:01.713 & $-$29:58:16.75 & 2.18$\times$10$^{-16}$ & 6.67$\times$10$^{-16}$ & 8.3\\
NGC~5236: XUV 15 & 13:37:06.354 & $-$29:57:51.15 & 1.97$\times$10$^{-16}$ & 1.97$\times$10$^{-16}$ & 7.9\\
NGC~5236: XUV 16 & 13:37:06.627 & $-$29:58:46.02 & 9.70$\times$10$^{-17}$ & 2.48$\times$10$^{-16}$ & 9.1\\
NGC~5236: XUV 17 & 13:37:09.362 & $-$29:59:08.94 & 9.12$\times$10$^{-17}$ & 1.42$\times$10$^{-16}$ & 9.7\\
NGC~5236: XUV 18 & 13:37:16.414 & $-$29:57:48.10 & 4.02$\times$10$^{-16}$ & 7.32$\times$10$^{-16}$ & 8.8\\
NGC~5236: XUV 19 & 13:37:06.730 & $-$29:57:42.46 & 4.74$\times$10$^{-17}$ & 8.60$\times$10$^{-17}$ & 7.7\\
NGC~5236: XUV 20$^{\dagger}$ & 13:37:02.316 & $-$29:57:18.89 & 1.80$\times$10$^{-15}$ & 7.10$\times$10$^{-15}$ & 7.0\\
NGC~5236: XUV 21$^{\dagger}$ & 13:37:11.409 & $-$29:55:34.79 & 3.24$\times$10$^{-15}$ & 1.18$\times$10$^{-14}$ & 5.6\\
NGC~5236: XUV 22$^{\dagger}$ & 13:37:07.142 & $-$29:56:47.37 & 1.24$\times$10$^{-15}$ & 3.00$\times$10$^{-15}$ & 6.6\\
\cutinhead{NGC~4625}
NGC~4625: XUV 01 & 12:42:07.985 & $+$41:14:13.65 & 1.33$\times$10$^{-16}$ & 3.75$\times$10$^{-16}$ & 10.0\\
NGC~4625: XUV 02 & 12:42:05.136 & $+$41:16:12.14 & 4.10$\times$10$^{-16}$ & 1.14$\times$10$^{-15}$ & 6.5\\
NGC~4625: XUV 03 & 12:42:04.025 & $+$41:16:35.98 & 1.73$\times$10$^{-16}$ & 3.71$\times$10$^{-16}$ & 5.9\\
NGC~4625: XUV 04 & 12:42:00.322 & $+$41:16:39.80 & 1.66$\times$10$^{-16}$ & 9.95$\times$10$^{-16}$ & 4.0\\
NGC~4625: XUV 05 & 12:41:59.150 & $+$41:15:21.11 & 2.05$\times$10$^{-16}$ & 3.20$\times$10$^{-16}$ & 4.5\\
NGC~4625: XUV 06 & 12:41:57.029 & $+$41:15:53.65 & 6.75$\times$10$^{-16}$ & 1.57$\times$10$^{-15}$ & 2.7\\
NGC~4625: XUV 07 & 12:41:56.234 & $+$41:15:30.90 & 4.66$\times$10$^{-16}$ & 1.21$\times$10$^{-15}$ & 3.1\\
NGC~4625: XUV 08 & 12:41:54.310 & $+$41:17:26.82 & 1.61$\times$10$^{-16}$ & 7.11$\times$10$^{-16}$ & 2.9\\
NGC~4625: XUV 09 & 12:41:51.538 & $+$41:15:17.52 & 1.02$\times$10$^{-16}$ & 2.78$\times$10$^{-16}$ & 3.2\\
NGC~4625: XUV 10 & 12:41:48.677 & $+$41:17:46.81 & 1.01$\times$10$^{-15}$ & 1.01$\times$10$^{-15}$ & 4.4\\
NGC~4625: XUV 11 & 12:41:45.151 & $+$41:16:00.29 & 2.60$\times$10$^{-16}$ & 2.35$\times$10$^{-15}$ & 4.3\\
NGC~4625: XUV 12 & 12:41:42.977 & $+$41:16:37.95 & 5.33$\times$10$^{-16}$ & 1.10$\times$10$^{-15}$ & 3.7\\
\enddata
\tablecomments{(1) Region identification, (2) Right Ascension, (3) Declination, (4) Observed H$\beta$ flux inside the corresponding slitlet. (5) Extinction-corrected H$\beta$ flux inside the slitlet; color excesses are given in Table~3. (6) Galactocentric distance in kpc. $^{\dagger}$Region belongs to the optical disk of M\,83.
}
\end{deluxetable}

\clearpage
\begin{deluxetable}{lrrrrrrrrrrr}
\rotate 
\setlength{\tabcolsep}{0.045in}
\tabletypesize{\tiny}
\tablecaption{XUV-regions line ratios\label{table2}}
\tablecolumns{12}
\tablewidth{0pt}
\tablehead{
\colhead{XUV region name} & \colhead{E($B-V$)} & \colhead{[OIII]/[OII]} & \colhead{[OIII]/H$\beta$} & \colhead{[NII]/H$\alpha$} & \colhead{[SII]/H$\alpha$} & \colhead{6717/6731} & \colhead{R23} & \colhead{log(EW$_{\mathrm{H}\alpha}$)} & \colhead{log(EW$_{\mathrm{[OIII]}}$)} & \colhead{Z$_{\mathrm{l/h,M91}}$} & \colhead{Z$_{\mathrm{l/h,PT05}}$} \\ 
\colhead{(1)} & \colhead{(2)} & \colhead{(3)} & \colhead{(4)} & \colhead{(5)}  & \colhead{(6)} & \colhead{(7)} & \colhead{(8)} & \colhead{(9)} & \colhead{(10)} & \colhead{(11)} & \colhead{(12)}
}
\startdata
\cutinhead{M\,83 (NGC~5236)}
NGC~5236: XUV 01 & 0.00 &   0.38 &   0.49 & $-$0.89  &   $-$1.11 & 1.79 &0.76&  2.71 &   2.28 &7.861/8.680 & 7.718/8.458 \\
NGC~5236: XUV 02 & 0.23 & $>$0.76 &   0.86 & $-$1.24  &  $<$$-$0.73 & \nodata  &$<$1.05&$>$1.80 & $>$1.91  & \nodata  & \nodata \\
NGC~5236: XUV 03 & 0.29 & $>$$-$0.20 &   0.14 & $-$0.65  &  $<$$-$0.47 & \nodata  &$<$0.68&  2.89 &   2.12 & \nodata   & \nodata \\
NGC~5236: XUV 04 & 0.00 &  \nodata & $<$$-$0.21 & $-$0.66  &  $<$$-$0.11 & \nodata  &$<$0.30&  2.23 & $<$2.00 & \nodata  & \nodata \\
NGC~5236: XUV 05 & 0.53 & $<$$-$0.42 & $<$$-$0.06 & $-$0.37  &   $-$0.50 & 1.04 &$<$0.63&  2.51 & $<$3.10 & \nodata  &  \nodata \\
NGC~5236: XUV 06 & 0.00 & $<$$-$0.52 & $<$0.01 & $-$0.51  &   $-$0.65 & 0.88 &$<$0.76&  2.96 & $<$1.53 & \nodata  &  \nodata \\
NGC~5236: XUV 07 & 0.44 &  $-$0.73 &  $-$0.07 & $-$0.60  &   $-$0.98 & 2.40 &0.85&  2.56 &   1.76 & 8.391/8.427 &  7.765/7.744 \\
NGC~5236: XUV 08 & 0.07 & $<$$-$0.78 & $<$0.12 & \nodata&  $<$$-$0.19 & \nodata  &$<$1.09&  2.75 & $<$1.64 & \nodata & \nodata \\
NGC~5236: XUV 09 & 0.12 &   0.06 &   0.39 & $-$0.89  &   $-$0.88 & 2.29 &0.79&  3.38 &   2.67 &8.015/8.617 & 7.891/8.316 \\
NGC~5236: XUV 10 & 0.10 &  $-$0.28 &   0.03 & $-$0.81  &   $-$0.76 & 1.11 &0.62&  2.99 &   2.10 &7.862/8.761 &  7.506/8.341 \\
NGC~5236: XUV 11 & 0.00 &   0.17 &   0.48 & $-$1.02  &   $-$1.11 & 2.01 &0.83&  2.52 &   2.42 &8.049/8.584 &  7.952/8.311\\
NGC~5236: XUV 12 & 0.49 &  $-$0.02 &   0.16 & $-$0.52  &   $-$0.84 & 1.07 &0.59&  2.72 &   2.34 &7.730/8.798 & 7.550/8.492\\
NGC~5236: XUV 13 & 0.19 &   0.04 &   0.32 & $-$0.69  &   $-$1.01 & 1.21 &0.72&  3.39 &   3.29 &7.904/8.689 & 7.751/8.391\\
NGC~5236: XUV 14 & 0.34 &  $-$0.29 &  $-$0.13 & $-$0.46  &   $-$0.74 & 0.87 &0.47&  2.63 &   1.63 &7.660/8.875 & 7.306/8.468 \\
NGC~5236: XUV 15 & 0.00 & $>$$-$0.56 &  $-$0.66 & $-$0.39  &   $-$0.60 & 0.77 &$<$0.14&  2.88 &   1.90 & \nodata   & \nodata \\
NGC~5236: XUV 16 & 0.28 & $<$$-$0.78 & $<$$-$0.46 & $-$0.34  &   $-$0.72 & 1.60 &$<$0.51&$>$1.91 & $<$0.81 & \nodata  & \nodata \\
NGC~5236: XUV 17 & 0.13 &  $-$0.58 &   0.05 & $-$0.85  &   $-$0.65 & 0.83 &0.86&  3.03 &   1.93 &8.348/8.442 & 7.877/7.824\\
NGC~5236: XUV 18 & 0.18 &  \nodata &   0.13 & $-$0.48  &   $-$0.93 & 1.06 &\nodata&  2.85 &   3.10 &  \nodata & \nodata \\
NGC~5236: XUV 19 & 0.18 &  \nodata & $<$0.03 & $-$0.43  &   $-$0.31 & 0.97 &$<$0.74&$>$1.41 & $<$1.84 & \nodata   & \nodata \\
NGC~5236: XUV 20 & 0.41 &  $-$0.49 &  $-$0.54 & $-$0.38  &  $<$$-$1.82 & \nodata  &0.20&  2.76 &   1.23 & \nodata/9.007 & \nodata/8.543 \\
NGC~5236: XUV 21 & 0.39 &  $-$0.29 &  $-$0.59 & $-$0.38  &   $-$0.93 & 1.31 &0.01&  3.05 &   1.78 & \nodata/9.061 & \nodata/8.696\\
NGC~5236: XUV 22 & 0.27 &   0.02 &  $-$0.27 & $-$0.38  &   $-$0.78 & 1.30 &0.14&$>$2.67 &   3.54 & \nodata/9.026 & \nodata/8.759 \\
\cutinhead{NGC~4625}
NGC~4625: XUV 01 & 0.31 &   0.51 &   0.59 &  $-$1.26 & $<$$-$0.59 &\nodata  &   0.83  &     2.85  &    2.54  &  7.941/\nodata  &  7.787/\nodata \\
NGC~4625: XUV 02 & 0.31 &  $-$0.43 &   0.30 &  $-$1.05 &  $-$0.94 &   1.90  &   0.99  &     2.76  &    2.24  &  8.531/\nodata  & 8.486/\nodata \\
NGC~4625: XUV 03 & 0.23 &\nodata & $<$$-$0.23 &\nodata &  $-$0.51 &   1.63  & $<$0.69  &     2.82  &  $<$2.28  &   \nodata   & \nodata \\ 
NGC~4625: XUV 04 & 0.54 & $<$$-$0.99 & $<$$-$0.23 &  $-$0.91 &  $-$0.63 &   1.19  & $<$0.92  &     2.77  &  $<$1.65  &   \nodata   & \nodata \\
NGC~4625: XUV 05 & 0.13 &  $-$0.25 &   0.22 &  $-$0.81 &  $-$0.79 &   0.90  &   0.79  &     2.37  &    2.00  &  8.123/\nodata  & 7.868/\nodata\\
NGC~4625: XUV 06 & 0.25 &  $-$0.65 &  $-$0.07 &  $-$0.81 &  $-$0.66 &   0.68  &   0.79  &     2.57  &    1.68  &  8.265/\nodata  & 7.635/\nodata  \\
NGC~4625: XUV 07 & 0.29 &  $-$0.09 &   0.44 &  $-$1.09 &  $-$1.04 &   1.20  &   0.92  &   $>$2.32  &  $>$2.07  &  8.296/\nodata &  8.269/\nodata \\
NGC~4625: XUV 08 & 0.45 &\nodata &   0.52 &  $-$0.99 & $<$$-$0.69 &\nodata  &   \nodata   &     3.22  &    2.63  &   \nodata   &  \nodata \\
NGC~4625: XUV 09 & 0.30 &   0.25 &   0.96 &  $-$0.85 & $<$$-$0.44 &\nodata  &   1.28  &     3.02  &    3.16  &  8.400/\nodata  & 8.400/\nodata  \\
NGC~4625: XUV 10 & 0.00 &\nodata &   0.68 &  $-$1.29 &  $-$0.66 &   1.36  &   \nodata   &     3.12  &    2.95  &   \nodata   &   \nodata \\
NGC~4625: XUV 11 & 0.66 & $>$0.26 &   0.92 &  $-$1.18 & $<$$-$0.68 &\nodata  & $<$1.24  &     2.61  &  $>$2.04  &   \nodata   &   \nodata \\ 
NGC~4625: XUV 12 & 0.22 &  $-$0.55 &   0.07 &  $-$1.10 &  $-$0.69 &   1.73  &   0.85  &     3.07  &    2.32  &  8.333/\nodata  &   7.891/\nodata 
\enddata
\tablecomments{(1) Region identification, (2) color excess derived from the H$\alpha$/H$\beta$ Balmer-line decrement; it includes the effects of both the Galactic and the internal dust extinction, (3) logarithm of the extinction-corrected [OIII]$\lambda$5007\AA/[OII]$\lambda\lambda$3726,3729\AA\AA\ line ratio, (4) logarithm of the [OIII]$\lambda$5007\AA/H$\beta$ line ratio, (5) logarithm of the [NII]$\lambda$6584\AA/H$\alpha$ line ratio, (6) logarithm of the [SII]$\lambda\lambda$6717,6731\AA\AA/H$\alpha$ line ratio, (7) [SII]$\lambda$6717\AA/[SII]$\lambda$6731\AA\ line ratio, (8) R23 parameter, defined as log\{([OII]$\lambda\lambda$3726,3729\AA\AA$+$[OIII]$\lambda$5007\AA$+$[OIII]$\lambda$4959\AA)/H$\beta$\}, (9) logarithm of the equivalent width (in angstroms) of H$\alpha$, (10) logarithm of the equivalent width of [OIII]$\lambda$5007\AA, (11) nebular oxygen abundance, 12+log(O/H), for the low/high-metallicity branch of the R23 calibration using the method of McGaugh (1991; M91), (12) oxygen abundances for the low/high-metallicity branch of the R23 calibration using the method of Pilyugin \& Thuan (2005; PT05). We have adopted 12+log(O/H)$_{\odot}$=8.69 throughout the text (Allende Prieto et al$.$ 2001).
}
\end{deluxetable}

\clearpage
\begin{deluxetable}{lccccccccc}
\rotate 
\tabletypesize{\footnotesize}
\tablecaption{Best-fitting single-star photoionization models (solar N/O ratio)}
\tablecolumns{10}
\tablewidth{0pt}
\tablehead{
\colhead{XUV region name} & \colhead{$\chi^2$}   & \colhead{f$_{\mathrm{H}\beta,\mathrm{model}}$} & \colhead{Age$_{*}$} & \colhead{T$_{\mathrm{eff},*}$} & \colhead{log(g)$_*$} & \colhead{Mass$_{*}$} & \colhead{Z$_{\mathrm{gas}}$} & \colhead{log(n$_{\mathrm{gas}}$)} & \colhead{log(R$_{\mathrm{in}}$)} \\
\colhead{}       &  \colhead{}        &  \colhead{(erg\,s$^{-1}$\,cm$^{-2}$)} &  \colhead{(Myr)} &            \colhead{(K)}      &      \colhead{}      & \colhead{(M$_{\odot}$)} & \colhead{(Z$_{\odot}$)} & \colhead{(cm$^{-3}$)} & \colhead{(cm)} \\
\colhead{(1)} & \colhead{(2)} & \colhead{(3)} & \colhead{(4)} & \colhead{(5)} & \colhead{(6)} & \colhead{(7)} & \colhead{(8)} &\colhead{(9)} & \colhead{(10)}
}
\startdata
\cutinhead{M\,83 (NGC~5236)}
NGC~5236: XUV 01 &   0.355 & 1.207E-15$_{6.160E-16}^{1.207E-15}$ & 3.0$_{3.0}^{3.0}$ & 36625$_{36625}^{36625}$ & 3.68$_{3.68}^{3.68}$ &  40.0$_{ 40.0}^{ 40.0}$ & 0.20$_{0.20}^{0.60}$ & 2.0$_{2.0}^{3.0}$ & 17.5$_{16.0}^{17.5}$ \\[2mm]
NGC~5236: XUV 07 &   1.975 & 3.196E-16$_{3.052E-16}^{3.203E-16}$ & 3.0$_{3.0}^{3.0}$ & 35636$_{35636}^{35636}$ & 3.96$_{3.96}^{3.96}$ &  25.0$_{ 25.0}^{ 25.0}$ & 0.20$_{0.20}^{0.30}$ & 2.0$_{2.0}^{2.0}$ & 16.0$_{16.0}^{17.5}$ \\[2mm]
NGC~5236: XUV 09 &   0.383 & 4.321E-16$_{4.117E-16}^{4.411E-15}$ & 1.0$_{1.0}^{3.0}$ & 37707$_{36625}^{40576}$ & 4.15$_{3.68}^{4.15}$ &  25.0$_{ 25.0}^{ 60.0}$ & 0.20$_{0.20}^{0.30}$ & 2.0$_{0.0}^{2.0}$ & 17.0$_{16.0}^{17.5}$ \\[2mm]
NGC~5236: XUV 10 &   0.413 & 1.379E-15$_{3.196E-16}^{1.379E-15}$ & 3.0$_{3.0}^{3.0}$ & 36625$_{35636}^{36625}$ & 3.68$_{3.68}^{3.96}$ &  40.0$_{ 25.0}^{ 40.0}$ & 0.10$_{0.10}^{0.20}$ & 0.0$_{0.0}^{2.0}$ & 17.5$_{16.0}^{17.5}$ \\[2mm]
NGC~5236: XUV 11 &   0.828 & 1.207E-15$_{1.201E-15}^{1.207E-15}$ & 3.0$_{3.0}^{3.0}$ & 36625$_{36625}^{36625}$ & 3.68$_{3.68}^{3.68}$ &  40.0$_{ 40.0}^{ 40.0}$ & 0.20$_{0.20}^{0.20}$ & 2.0$_{2.0}^{2.0}$ & 17.5$_{16.0}^{17.5}$ \\[2mm]
NGC~5236: XUV 12 &   0.427 & 2.592E-16$_{2.385E-16}^{2.733E-16}$ & 3.0$_{3.0}^{3.0}$ & 35636$_{35636}^{35636}$ & 3.96$_{3.96}^{3.96}$ &  25.0$_{ 25.0}^{ 25.0}$ & 0.60$_{0.40}^{0.80}$ & 3.0$_{3.0}^{3.0}$ & 17.5$_{17.5}^{17.5}$ \\[2mm]
NGC~5236: XUV 13 &   0.207 & 2.677E-16$_{2.326E-16}^{2.901E-16}$ & 3.0$_{3.0}^{3.0}$ & 35636$_{35636}^{35636}$ & 3.96$_{3.96}^{3.96}$ &  25.0$_{ 25.0}^{ 25.0}$ & 0.30$_{0.20}^{0.50}$ & 3.0$_{3.0}^{3.0}$ & 17.0$_{16.0}^{17.5}$ \\[2mm]
NGC~5236: XUV 14 &   0.281 & 1.654E-16$_{1.385E-16}^{1.654E-16}$ & 1.0$_{1.0}^{1.0}$ & 34984$_{34984}^{34984}$ & 4.18$_{4.18}^{4.18}$ &  20.0$_{ 20.0}^{ 20.0}$ & 0.80$_{0.80}^{1.00}$ & 2.0$_{2.0}^{2.0}$ & 16.0$_{16.0}^{16.0}$ \\[2mm]
NGC~5236: XUV 17 &   1.848 & 1.351E-15$_{1.351E-15}^{1.351E-15}$ & 3.0$_{3.0}^{3.0}$ & 36625$_{36625}^{36625}$ & 3.68$_{3.68}^{3.68}$ &  40.0$_{ 40.0}^{ 40.0}$ & 0.20$_{0.20}^{0.20}$ & 0.0$_{0.0}^{0.0}$ & 16.5$_{16.0}^{17.5}$ \\[2mm]
NGC~5236: XUV 20 &   0.419 & 1.044E-16$_{6.723E-17}^{1.044E-16}$ & 1.0$_{1.0}^{3.0}$ & 34984$_{33751}^{34984}$ & 4.18$_{4.05}^{4.18}$ &  20.0$_{ 20.0}^{ 20.0}$ & 1.00$_{1.00}^{1.50}$ & 3.0$_{3.0}^{3.0}$ & 17.5$_{16.0}^{17.5}$ \\[2mm]
NGC~5236: XUV 21 &   0.741 & 9.366E-17$_{5.452E-17}^{9.366E-17}$ & 1.0$_{1.0}^{3.0}$ & 34984$_{33751}^{34984}$ & 4.18$_{4.05}^{4.18}$ &  20.0$_{ 20.0}^{ 20.0}$ & 1.50$_{1.50}^{2.00}$ & 3.0$_{3.0}^{3.0}$ & 17.5$_{16.0}^{17.5}$ \\[2mm]
NGC~5236: XUV 22 &   1.486 & 6.525E-17$_{6.362E-17}^{7.187E-17}$ & 1.0$_{1.0}^{1.0}$ & 34984$_{34984}^{34984}$ & 4.18$_{4.18}^{4.18}$ &  20.0$_{ 20.0}^{ 20.0}$ & 1.50$_{1.50}^{1.50}$ & 3.0$_{3.0}^{3.0}$ & 16.5$_{16.0}^{17.0}$ \\[2mm]
\cutinhead{NGC~4625}
NGC~4625: XUV 01 &   0.702 & 1.126E-15$_{2.278E-16}^{4.545E-15}$ & 1.0$_{1.0}^{3.0}$ & 45087$_{36625}^{49565}$ & 3.99$_{3.68}^{3.99}$ &  60.0$_{ 40.0}^{120.0}$ & 0.10$_{0.10}^{0.20}$ & 1.0$_{0.0}^{3.0}$ & 16.0$_{16.0}^{17.5}$ \\[2mm]
NGC~4625: XUV 02 &   4.260 & 4.640E-16$_{3.031E-16}^{1.020E-15}$ & 1.0$_{1.0}^{3.0}$ & 42600$_{36625}^{42600}$ & 4.08$_{3.68}^{4.08}$ &  40.0$_{ 40.0}^{ 60.0}$ & 0.10$_{0.10}^{0.10}$ & 0.0$_{0.0}^{1.0}$ & 16.5$_{16.0}^{17.5}$ \\[2mm]
NGC~4625: XUV 05 &   0.609 & 2.914E-16$_{1.025E-16}^{2.921E-16}$ & 3.0$_{1.0}^{3.0}$ & 36625$_{36625}^{37707}$ & 3.68$_{3.68}^{4.15}$ &  40.0$_{ 25.0}^{ 40.0}$ & 0.20$_{0.20}^{0.30}$ & 1.0$_{1.0}^{1.0}$ & 16.0$_{16.0}^{17.5}$ \\[2mm]
NGC~4625: XUV 06 &   1.478 & 3.031E-16$_{1.051E-16}^{3.094E-16}$ & 3.0$_{1.0}^{3.0}$ & 36625$_{36625}^{37707}$ & 3.68$_{3.68}^{4.15}$ &  40.0$_{ 25.0}^{ 40.0}$ & 0.20$_{0.10}^{0.20}$ & 0.0$_{0.0}^{0.0}$ & 17.0$_{16.0}^{17.5}$ \\[2mm]
NGC~4625: XUV 07 &   2.504 & 2.695E-16$_{2.695E-16}^{1.158E-15}$ & 3.0$_{1.0}^{3.0}$ & 36625$_{36625}^{45087}$ & 3.68$_{3.68}^{3.99}$ &  40.0$_{ 40.0}^{ 60.0}$ & 0.20$_{0.10}^{0.20}$ & 2.0$_{0.0}^{2.0}$ & 16.5$_{16.0}^{17.5}$ \\[2mm]
NGC~4625: XUV 09 &   5.162 & 3.218E-15$_{3.218E-15}^{4.193E-15}$ & 1.0$_{1.0}^{1.0}$ & 49565$_{49565}^{49565}$ & 3.92$_{3.92}^{3.92}$ & 120.0$_{120.0}^{120.0}$ & 0.60$_{0.30}^{0.60}$ & 1.0$_{0.0}^{1.0}$ & 16.0$_{16.0}^{17.5}$ \\[2mm]
NGC~4625: XUV 12 &   3.405 & 1.056E-16$_{1.053E-16}^{3.031E-16}$ & 1.0$_{1.0}^{3.0}$ & 37707$_{36625}^{37707}$ & 4.15$_{3.68}^{4.15}$ &  25.0$_{ 25.0}^{ 40.0}$ & 0.10$_{0.10}^{0.10}$ & 1.0$_{1.0}^{1.0}$ & 17.0$_{16.0}^{17.5}$ \\[2mm]
\enddata
\tablecomments{(1) Region identification, (2) best-fitting reduced $\chi^2$, (3) H$\beta$ flux predicted by the photoionization models, (4) age of the star (Myr), (5) effective temperature (in K), (6) logarithm of the gravity, (7) ZAMS mass (M$_{\odot}$) , (8) metal abundance of the gas and dust grains (in Z$_{\odot}$), (9) density of the gas (in atoms\, cm$^{-3}$), (10) inner radius of the spherical gas distribution (in cm). The 68.3\% confidence intervals for each parameter considered separately are also given.
}
\end{deluxetable}

\clearpage
\begin{deluxetable}{lccccccccc}
\rotate 
\tabletypesize{\footnotesize}
\tablecaption{Best-fitting single-star photoionization models (O/H-scaled N/O ratio)}
\tablecolumns{10}
\tablewidth{0pt}
\tablehead{
\colhead{XUV region name} & \colhead{$\chi^2$}   & \colhead{f$_{\mathrm{H}\beta,\mathrm{model}}$} & \colhead{Age$_{*}$} & \colhead{T$_{\mathrm{eff},*}$} & \colhead{log(g)$_*$} & \colhead{Mass$_{*}$} & \colhead{Z$_{\mathrm{gas}}$} & \colhead{log(n$_{\mathrm{gas}}$)} & \colhead{log(R$_{\mathrm{in}}$)} \\
\colhead{}       &  \colhead{}        &  \colhead{(erg\,s$^{-1}$\,cm$^{-2}$)} &  \colhead{(Myr)} &            \colhead{(K)}      &      \colhead{}      & \colhead{(M$_{\odot}$)} & \colhead{(Z$_{\odot}$)} & \colhead{(cm$^{-3}$)} & \colhead{(cm)} \\
\colhead{(1)} & \colhead{(2)} & \colhead{(3)} & \colhead{(4)} & \colhead{(5)} & \colhead{(6)} & \colhead{(7)} & \colhead{(8)} &\colhead{(9)} & \colhead{(10)}
}
\startdata
\cutinhead{M\,83 (NGC~5236)}
NGC~5236: XUV 01 &   3.022 & 7.773E-16$_{1.269E-16}^{8.290E-15}$ & 1.0$_{1.0}^{1.0}$ & 42600$_{37707}^{49565}$ & 4.08$_{3.92}^{4.15}$ &  40.0$_{ 25.0}^{120.0}$ & 2.00$_{2.00}^{2.00}$ & 2.0$_{1.0}^{3.0}$ & 16.0$_{16.0}^{17.5}$ \\[2mm]
NGC~5236: XUV 07 &   5.030 & 1.537E-16$_{1.537E-16}^{1.537E-16}$ & 1.0$_{1.0}^{1.0}$ & 34984$_{34984}^{34984}$ & 4.18$_{4.18}^{4.18}$ &  20.0$_{ 20.0}^{ 20.0}$ & 1.00$_{1.00}^{1.00}$ & 2.0$_{2.0}^{2.0}$ & 16.0$_{16.0}^{16.0}$ \\[2mm]
NGC~5236: XUV 09 &   2.892 & 1.424E-14$_{1.716E-16}^{1.424E-14}$ & 1.0$_{1.0}^{3.0}$ & 49565$_{35636}^{49565}$ & 3.92$_{3.92}^{3.96}$ & 120.0$_{ 25.0}^{120.0}$ & 1.50$_{1.00}^{1.50}$ & 0.0$_{0.0}^{3.0}$ & 17.0$_{16.0}^{17.5}$ \\[2mm]
NGC~5236: XUV 10 &   0.905 & 1.685E-16$_{1.685E-16}^{1.685E-16}$ & 1.0$_{1.0}^{1.0}$ & 34984$_{34984}^{34984}$ & 4.18$_{4.18}^{4.18}$ &  20.0$_{ 20.0}^{ 20.0}$ & 0.80$_{0.80}^{0.80}$ & 2.0$_{2.0}^{2.0}$ & 16.0$_{16.0}^{16.0}$ \\[2mm]
NGC~5236: XUV 11 &   2.589 & 1.716E-16$_{1.681E-16}^{2.026E-16}$ & 3.0$_{3.0}^{3.0}$ & 35636$_{35636}^{35636}$ & 3.96$_{3.96}^{3.96}$ &  25.0$_{ 25.0}^{ 25.0}$ & 1.00$_{0.80}^{1.00}$ & 3.0$_{3.0}^{3.0}$ & 16.5$_{16.0}^{17.0}$ \\[2mm]
NGC~5236: XUV 12 &   6.731 & 1.685E-16$_{1.537E-16}^{1.685E-16}$ & 1.0$_{1.0}^{1.0}$ & 34984$_{34984}^{34984}$ & 4.18$_{4.18}^{4.18}$ &  20.0$_{ 20.0}^{ 20.0}$ & 0.80$_{0.80}^{1.00}$ & 2.0$_{2.0}^{2.0}$ & 16.0$_{16.0}^{16.0}$ \\[2mm]
NGC~5236: XUV 13 &   5.489 & 1.461E-16$_{1.296E-16}^{1.424E-14}$ & 3.0$_{1.0}^{3.0}$ & 35636$_{35636}^{49565}$ & 3.96$_{3.92}^{3.96}$ &  25.0$_{ 25.0}^{120.0}$ & 1.50$_{1.00}^{1.50}$ & 3.0$_{0.0}^{3.0}$ & 17.0$_{16.0}^{17.5}$ \\[2mm]
NGC~5236: XUV 14 &   4.918 & 6.495E-17$_{6.347E-17}^{1.537E-16}$ & 1.0$_{1.0}^{1.0}$ & 34984$_{34984}^{34984}$ & 4.18$_{4.18}^{4.18}$ &  20.0$_{ 20.0}^{ 20.0}$ & 1.50$_{1.00}^{1.50}$ & 3.0$_{2.0}^{3.0}$ & 16.5$_{16.0}^{16.5}$ \\[2mm]
NGC~5236: XUV 17 &   1.338 & 1.685E-16$_{1.685E-16}^{1.685E-16}$ & 1.0$_{1.0}^{1.0}$ & 34984$_{34984}^{34984}$ & 4.18$_{4.18}^{4.18}$ &  20.0$_{ 20.0}^{ 20.0}$ & 0.80$_{0.80}^{0.80}$ & 2.0$_{2.0}^{2.0}$ & 16.0$_{16.0}^{16.0}$ \\[2mm]
NGC~5236: XUV 20 &   5.418 & 6.160E-17$_{4.705E-17}^{9.323E-17}$ & 1.0$_{1.0}^{3.0}$ & 34984$_{33751}^{34984}$ & 4.18$_{4.05}^{4.18}$ &  20.0$_{ 20.0}^{ 20.0}$ & 2.00$_{1.50}^{2.00}$ & 3.0$_{3.0}^{3.0}$ & 17.0$_{16.0}^{17.5}$ \\[2mm]
NGC~5236: XUV 21 &   3.836 & 4.705E-17$_{4.545E-17}^{8.523E-17}$ & 3.0$_{1.0}^{3.0}$ & 33751$_{33751}^{34984}$ & 4.05$_{4.05}^{4.18}$ &  20.0$_{ 20.0}^{ 20.0}$ & 2.00$_{2.00}^{2.00}$ & 3.0$_{3.0}^{3.0}$ & 16.5$_{16.0}^{17.5}$ \\[2mm]
NGC~5236: XUV 22 &   4.836 & 5.465E-17$_{5.267E-17}^{6.160E-17}$ & 1.0$_{1.0}^{1.0}$ & 34984$_{34984}^{34984}$ & 4.18$_{4.18}^{4.18}$ &  20.0$_{ 20.0}^{ 20.0}$ & 2.00$_{2.00}^{2.00}$ & 3.0$_{3.0}^{3.0}$ & 16.5$_{16.0}^{17.0}$ \\[2mm]
\cutinhead{NGC~4625}
NGC~4625: XUV 01 &   0.117 & 7.990E-16$_{7.972E-16}^{2.170E-15}$ & 1.0$_{1.0}^{2.0}$ & 45087$_{45087}^{47548}$ & 3.99$_{3.73}^{3.99}$ &  60.0$_{ 60.0}^{120.0}$ & 1.00$_{1.00}^{1.50}$ & 1.0$_{1.0}^{1.0}$ & 17.5$_{16.0}^{17.5}$ \\[2mm]
NGC~4625: XUV 02 &   4.120 & 5.735E-17$_{4.815E-17}^{6.117E-17}$ & 5.0$_{1.0}^{5.0}$ & 31698$_{31698}^{35636}$ & 3.65$_{3.65}^{4.18}$ &  25.0$_{ 20.0}^{ 25.0}$ & 0.50$_{0.40}^{0.80}$ & 1.0$_{1.0}^{3.0}$ & 16.0$_{16.0}^{17.5}$ \\[2mm]
NGC~4625: XUV 05 &   2.504 & 3.781E-17$_{3.781E-17}^{4.815E-17}$ & 1.0$_{1.0}^{1.0}$ & 34984$_{34984}^{34984}$ & 4.18$_{4.18}^{4.18}$ &  20.0$_{ 20.0}^{ 20.0}$ & 0.80$_{0.80}^{0.80}$ & 2.0$_{1.0}^{2.0}$ & 16.0$_{16.0}^{16.5}$ \\[2mm]
NGC~4625: XUV 06 &   1.256 & 3.781E-17$_{3.781E-17}^{3.781E-17}$ & 1.0$_{1.0}^{1.0}$ & 34984$_{34984}^{34984}$ & 4.18$_{4.18}^{4.18}$ &  20.0$_{ 20.0}^{ 20.0}$ & 0.80$_{0.80}^{0.80}$ & 2.0$_{2.0}^{2.0}$ & 16.0$_{16.0}^{16.0}$ \\[2mm]
NGC~4625: XUV 07 &   2.333 & 4.927E-17$_{4.927E-17}^{4.927E-17}$ & 1.0$_{1.0}^{1.0}$ & 34984$_{34984}^{34984}$ & 4.18$_{4.18}^{4.18}$ &  20.0$_{ 20.0}^{ 20.0}$ & 0.50$_{0.50}^{0.50}$ & 2.0$_{2.0}^{2.0}$ & 16.0$_{16.0}^{16.0}$ \\[2mm]
NGC~4625: XUV 09 &   0.280 & 1.839E-16$_{1.839E-16}^{1.839E-16}$ & 1.0$_{1.0}^{1.0}$ & 34984$_{34984}^{34984}$ & 4.18$_{4.18}^{4.18}$ &  20.0$_{ 20.0}^{ 20.0}$ & 0.80$_{0.80}^{0.80}$ & 0.0$_{0.0}^{0.0}$ & 16.5$_{16.5}^{16.5}$ \\[2mm]
NGC~4625: XUV 12 &   1.366 & 3.781E-17$_{3.781E-17}^{3.781E-17}$ & 1.0$_{1.0}^{1.0}$ & 34984$_{34984}^{34984}$ & 4.18$_{4.18}^{4.18}$ &  20.0$_{ 20.0}^{ 20.0}$ & 0.80$_{0.80}^{0.80}$ & 2.0$_{2.0}^{2.0}$ & 16.0$_{16.0}^{16.0}$ \\[2mm]
\enddata
\tablecomments{(1) Region identification, (2) best-fitting reduced $\chi^2$, (3) H$\beta$ flux predicted by the photoionization models, (4) age of the star (Myr), (5) effective temperature (in K), (6) logarithm of the gravity, (7) ZAMS mass (M$_{\odot}$) , (8) metal abundance of the gas and dust grains (in Z$_{\odot}$), (9) density of the gas (in atoms\, cm$^{-3}$), (10) inner radius of the spherical gas distribution (in cm). The 68.3\% confidence intervals for each parameter considered separately are also given.
}
\end{deluxetable}

\begin{deluxetable}{lcccccc}
\tablecaption{Best-fitting starburst photoionization models (solar N/O ratio)}
\tablecolumns{7}
\tablewidth{0pt}
\tablehead{
\colhead{XUV region name} & \colhead{$\chi^2$}   & \colhead{f$_{\mathrm{H}\beta,\mathrm{model}}$} & \colhead{Age$_{\mathrm{starburst}}$} & \colhead{Z$_{\mathrm{gas}}$} & \colhead{n$_{\mathrm{gas}}$} & \colhead{log(R$_{\mathrm{in}}$)} \\
\colhead{}       &  \colhead{}        &  \colhead{(erg\,s$^{-1}$\,cm$^{-2}$)} &  \colhead{(Myr)} & \colhead{(Z$_{\odot}$)} & \colhead{(cm$^{-3}$)} & \colhead{(cm)} \\
\colhead{(1)} & \colhead{(2)} & \colhead{(3)} & \colhead{(4)} & \colhead{(5)} &\colhead{(6)} & \colhead{(7)}
}
\startdata
\cutinhead{M\,83 (NGC~5236)}
NGC~5236: XUV 01 &   1.845 & 7.972E-13$_{7.972E-13}^{1.022E-12}$ & 5.0$_{5.0}^{5.0}$ & 0.20$_{0.20}^{0.20}$ & 1.5$_{1.0}^{1.5}$ & 16.5$_{16.0}^{17.5}$ \\[2mm]
NGC~5236: XUV 07 &   2.487 & 6.075E-13$_{6.075E-13}^{7.171E-13}$ & 6.0$_{5.0}^{6.0}$ & 0.20$_{0.20}^{0.30}$ & 1.0$_{0.5}^{1.0}$ & 16.5$_{16.0}^{17.5}$ \\[2mm]
NGC~5236: XUV 09 &   2.471 & 1.022E-12$_{1.022E-12}^{1.238E-12}$ & 5.0$_{5.0}^{5.0}$ & 0.20$_{0.20}^{0.20}$ & 1.0$_{0.5}^{1.0}$ & 16.0$_{16.0}^{17.5}$ \\[2mm]
NGC~5236: XUV 10 &   2.414 & 1.238E-12$_{2.515E-13}^{1.238E-12}$ & 5.0$_{5.0}^{5.0}$ & 0.20$_{0.20}^{0.60}$ & 0.5$_{0.5}^{1.5}$ & 16.5$_{16.0}^{17.5}$ \\[2mm]
NGC~5236: XUV 11 &   1.776 & 8.425E-14$_{8.425E-14}^{7.972E-13}$ & 2.0$_{2.0}^{5.0}$ & 1.50$_{0.20}^{1.50}$ & 2.0$_{1.5}^{2.0}$ & 16.0$_{16.0}^{17.5}$ \\[2mm]
NGC~5236: XUV 12 &   3.387 & 1.238E-12$_{3.695E-13}^{1.238E-12}$ & 5.0$_{5.0}^{5.0}$ & 0.20$_{0.20}^{0.40}$ & 0.5$_{0.5}^{1.5}$ & 17.0$_{16.0}^{17.5}$ \\[2mm]
NGC~5236: XUV 13 &   1.372 & 1.238E-12$_{1.238E-12}^{1.238E-12}$ & 5.0$_{5.0}^{5.0}$ & 0.20$_{0.20}^{0.20}$ & 0.5$_{0.5}^{0.5}$ & 16.0$_{16.0}^{17.5}$ \\[2mm]
NGC~5236: XUV 14 &   2.488 & 6.754E-13$_{6.754E-13}^{7.899E-13}$ & 5.0$_{5.0}^{5.0}$ & 0.40$_{0.30}^{0.40}$ & 0.5$_{0.5}^{0.5}$ & 17.5$_{16.0}^{17.5}$ \\[2mm]
NGC~5236: XUV 17 &   4.780 & 1.238E-12$_{1.238E-12}^{1.238E-12}$ & 5.0$_{5.0}^{5.0}$ & 0.20$_{0.20}^{0.20}$ & 0.5$_{0.5}^{0.5}$ & 17.0$_{16.0}^{17.5}$ \\[2mm]
NGC~5236: XUV 20 &   2.350 & 1.565E-13$_{7.287E-15}^{1.852E-13}$ & 6.0$_{6.0}^{6.0}$ & 0.60$_{0.50}^{1.00}$ & 1.5$_{1.5}^{3.5}$ & 16.5$_{16.0}^{17.5}$ \\[2mm]
NGC~5236: XUV 21 &   4.106 & 1.572E-13$_{1.565E-13}^{1.572E-13}$ & 6.0$_{6.0}^{6.0}$ & 0.60$_{0.60}^{0.60}$ & 1.5$_{1.5}^{1.5}$ & 17.5$_{16.0}^{17.5}$ \\[2mm]
NGC~5236: XUV 22 &   4.888 & 5.207E-13$_{5.207E-13}^{5.896E-13}$ & 5.0$_{5.0}^{5.0}$ & 0.60$_{0.50}^{0.60}$ & 0.5$_{0.5}^{0.5}$ & 17.0$_{16.0}^{17.5}$ \\[2mm]
\cutinhead{NGC~4625}
NGC~4625: XUV 01 &   0.519 & 1.020E-14$_{2.374E-15}^{3.116E-13}$ & 2.0$_{2.0}^{5.0}$ & 1.50$_{0.20}^{2.00}$ & 2.5$_{0.5}^{3.5}$ & 17.5$_{16.0}^{17.5}$ \\[2mm]
NGC~4625: XUV 02 &   5.265 & 1.890E-14$_{1.890E-14}^{2.777E-13}$ & 2.0$_{2.0}^{5.0}$ & 1.50$_{0.20}^{1.50}$ & 2.0$_{0.5}^{2.0}$ & 16.0$_{16.0}^{17.5}$ \\[2mm]
NGC~4625: XUV 05 &   2.255 & 2.777E-13$_{2.777E-13}^{2.777E-13}$ & 5.0$_{5.0}^{5.0}$ & 0.20$_{0.20}^{0.20}$ & 0.5$_{0.5}^{0.5}$ & 16.0$_{16.0}^{17.5}$ \\[2mm]
NGC~4625: XUV 06 &   4.145 & 1.405E-13$_{1.158E-13}^{1.772E-13}$ & 5.0$_{5.0}^{5.0}$ & 0.30$_{0.30}^{0.40}$ & 1.0$_{0.5}^{1.0}$ & 16.0$_{16.0}^{17.5}$ \\[2mm]
NGC~4625: XUV 07 &   2.453 & 1.890E-14$_{1.890E-14}^{1.939E-14}$ & 2.0$_{2.0}^{2.0}$ & 1.50$_{1.50}^{1.50}$ & 2.0$_{2.0}^{2.0}$ & 16.0$_{16.0}^{17.5}$ \\[2mm]
NGC~4625: XUV 09 &   8.609 & 2.758E-15$_{2.758E-15}^{2.803E-15}$ & 2.0$_{2.0}^{2.0}$ & 2.00$_{2.00}^{2.00}$ & 3.0$_{3.0}^{3.0}$ & 16.0$_{16.0}^{16.5}$ \\[2mm]
NGC~4625: XUV 12 &   6.202 & 3.644E-14$_{3.644E-14}^{3.472E-13}$ & 2.0$_{2.0}^{5.0}$ & 1.50$_{0.10}^{1.50}$ & 1.5$_{0.5}^{1.5}$ & 16.5$_{16.0}^{17.5}$ \\[2mm]
\enddata
\tablecomments{(1) Region identification, (2) best-fitting reduced $\chi^2$, (3) H$\beta$ flux predicted by the photoionization models, (4) age of the starburst (Myr), (5) metal abundance of the gas and dust grains (in Z$_{\odot}$), (6) density of the gas (in cm$^{-3}$), (7) inner radius of the spherical gas distribution (in cm). A stellar mass of 10$^{3}$\,M$_{\odot}$ was adopted for the ionizing starburst. The 68.3\% confidence intervals for each parameter considered separately are also given (see text for more details).
}
\end{deluxetable}

\begin{deluxetable}{lcccccc}
\tablecaption{Best-fitting starburst photoionization models (O/H-scaled N/O ratio)}
\tablecolumns{7}
\tablewidth{0pt}
\tablehead{
\colhead{XUV region name} & \colhead{$\chi^2$}   & \colhead{f$_{\mathrm{H}\beta,\mathrm{model}}$} & \colhead{Age$_{\mathrm{starburst}}$} & \colhead{Z$_{\mathrm{gas}}$} & \colhead{n$_{\mathrm{gas}}$} & \colhead{log(R$_{\mathrm{in}}$)} \\
\colhead{}       &  \colhead{}        &  \colhead{(erg\,s$^{-1}$\,cm$^{-2}$)} &  \colhead{(Myr)} & \colhead{(Z$_{\odot}$)} & \colhead{(cm$^{-3}$)} & \colhead{(cm)} \\
\colhead{(1)} & \colhead{(2)} & \colhead{(3)} & \colhead{(4)} & \colhead{(5)} &\colhead{(6)} & \colhead{(7)}
}
\startdata
\cutinhead{M\,83 (NGC~5236)}
NGC~5236: XUV 01 &   3.148 & 5.316E-14$_{5.316E-14}^{5.490E-14}$ & 2.0$_{2.0}^{2.0}$ & 2.00$_{2.00}^{2.00}$ & 2.0$_{2.0}^{2.0}$ & 16.0$_{16.0}^{17.5}$ \\[2mm]
NGC~5236: XUV 07 &  15.092 & 7.254E-15$_{7.254E-15}^{7.254E-15}$ & 6.0$_{6.0}^{6.0}$ & 1.00$_{1.00}^{1.00}$ & 3.5$_{3.5}^{3.5}$ & 16.0$_{16.0}^{16.0}$ \\[2mm]
NGC~5236: XUV 09 &   4.676 & 5.316E-14$_{5.316E-14}^{5.490E-14}$ & 2.0$_{2.0}^{2.0}$ & 2.00$_{2.00}^{2.00}$ & 2.0$_{2.0}^{2.0}$ & 16.0$_{16.0}^{17.5}$ \\[2mm]
NGC~5236: XUV 10 &   8.301 & 1.066E-13$_{1.063E-13}^{1.073E-13}$ & 2.0$_{2.0}^{2.0}$ & 2.00$_{2.00}^{2.00}$ & 1.5$_{1.5}^{1.5}$ & 17.0$_{16.0}^{17.5}$ \\[2mm]
NGC~5236: XUV 11 &   1.969 & 5.316E-14$_{5.316E-14}^{5.490E-14}$ & 2.0$_{2.0}^{2.0}$ & 2.00$_{2.00}^{2.00}$ & 2.0$_{2.0}^{2.0}$ & 16.0$_{16.0}^{17.5}$ \\[2mm]
NGC~5236: XUV 12 &  14.941 & 5.316E-14$_{5.316E-14}^{5.490E-14}$ & 2.0$_{2.0}^{2.0}$ & 2.00$_{2.00}^{2.00}$ & 2.0$_{2.0}^{2.0}$ & 16.0$_{16.0}^{17.5}$ \\[2mm]
NGC~5236: XUV 13 &   7.870 & 5.316E-14$_{5.316E-14}^{5.490E-14}$ & 2.0$_{2.0}^{2.0}$ & 2.00$_{2.00}^{2.00}$ & 2.0$_{2.0}^{2.0}$ & 16.0$_{16.0}^{17.5}$ \\[2mm]
NGC~5236: XUV 14 &  15.620 & 3.038E-15$_{3.038E-15}^{3.038E-15}$ & 5.0$_{5.0}^{5.0}$ & 2.00$_{2.00}^{2.00}$ & 3.5$_{3.5}^{3.5}$ & 17.0$_{17.0}^{17.0}$ \\[2mm]
NGC~5236: XUV 17 &  11.446 & 1.613E-13$_{1.613E-13}^{1.628E-13}$ & 2.0$_{2.0}^{2.0}$ & 1.50$_{1.50}^{1.50}$ & 1.5$_{1.5}^{1.5}$ & 16.0$_{16.0}^{17.5}$ \\[2mm]
NGC~5236: XUV 20 &   3.247 & 3.038E-15$_{3.038E-15}^{3.038E-15}$ & 5.0$_{5.0}^{5.0}$ & 2.00$_{2.00}^{2.00}$ & 3.5$_{3.5}^{3.5}$ & 17.0$_{17.0}^{17.0}$ \\[2mm]
NGC~5236: XUV 21 &   7.273 & 3.038E-15$_{2.441E-15}^{3.038E-15}$ & 5.0$_{5.0}^{5.0}$ & 2.00$_{2.00}^{2.00}$ & 3.5$_{3.5}^{3.5}$ & 17.0$_{16.0}^{17.0}$ \\[2mm]
NGC~5236: XUV 22 &  14.128 & 3.038E-15$_{3.038E-15}^{3.038E-15}$ & 5.0$_{5.0}^{5.0}$ & 2.00$_{2.00}^{2.00}$ & 3.5$_{3.5}^{3.5}$ & 17.0$_{17.0}^{17.0}$ \\[2mm]
\cutinhead{NGC~4625}
NGC~4625: XUV 01 &   0.640 & 6.347E-15$_{5.802E-15}^{6.347E-15}$ & 2.0$_{2.0}^{2.0}$ & 2.00$_{2.00}^{2.00}$ & 2.5$_{2.5}^{2.5}$ & 17.5$_{16.0}^{17.5}$ \\[2mm]
NGC~4625: XUV 02 &   6.409 & 1.193E-14$_{1.193E-14}^{1.232E-14}$ & 2.0$_{2.0}^{2.0}$ & 2.00$_{2.00}^{2.00}$ & 2.0$_{2.0}^{2.0}$ & 16.0$_{16.0}^{17.5}$ \\[2mm]
NGC~4625: XUV 05 &   8.382 & 1.193E-14$_{1.193E-14}^{1.232E-14}$ & 2.0$_{2.0}^{2.0}$ & 2.00$_{2.00}^{2.00}$ & 2.0$_{2.0}^{2.0}$ & 16.0$_{16.0}^{17.5}$ \\[2mm]
NGC~4625: XUV 06 &  12.131 & 2.385E-14$_{2.385E-14}^{6.646E-14}$ & 2.0$_{2.0}^{2.0}$ & 2.00$_{1.50}^{2.00}$ & 1.5$_{1.0}^{1.5}$ & 16.0$_{16.0}^{17.5}$ \\[2mm]
NGC~4625: XUV 07 &   3.150 & 1.193E-14$_{1.193E-14}^{1.232E-14}$ & 2.0$_{2.0}^{2.0}$ & 2.00$_{2.00}^{2.00}$ & 2.0$_{2.0}^{2.0}$ & 16.0$_{16.0}^{17.5}$ \\[2mm]
NGC~4625: XUV 09 &   7.502 & 2.758E-15$_{2.758E-15}^{2.803E-15}$ & 2.0$_{2.0}^{2.0}$ & 2.00$_{2.00}^{2.00}$ & 3.0$_{3.0}^{3.0}$ & 16.0$_{16.0}^{16.5}$ \\[2mm]
NGC~4625: XUV 12 &   6.645 & 3.619E-14$_{3.619E-14}^{3.652E-14}$ & 2.0$_{2.0}^{2.0}$ & 1.50$_{1.50}^{1.50}$ & 1.5$_{1.5}^{1.5}$ & 16.0$_{16.0}^{17.5}$ \\[2mm]
\enddata
\tablecomments{(1) Region identification, (2) best-fitting reduced $\chi^2$, (3) H$\beta$ flux predicted by the photoionization models, (4) age of the starburst (Myr), (5) metal abundance of the gas and dust grains (in Z$_{\odot}$), (6) density of the gas (in cm$^{-3}$), (7) inner radius of the spherical gas distribution (in cm). A stellar mass of 10$^{3}$\,M$_{\odot}$ was adopted for the ionizing starburst. The 68.3\% confidence intervals for each parameter considered separately are also given (see text for more details).
}
\end{deluxetable}

\begin{deluxetable}{lcccc|cccc|cccc|cccc|cccc}
\rotate 
\setlength{\tabcolsep}{0.019in}
\tabletypesize{\tiny}
\tablecaption{Observed and best-fitting line ratios}
\tablecolumns{21}
\tablewidth{0pt}
\tablehead{
\colhead{XUV region name} & \multicolumn{4}{c}{Observed line ratios} &  \multicolumn{4}{c}{Single-star, solar-N/O best fit} & \multicolumn{4}{c}{Single-star, O/H-scaled-N/O best fit} & \multicolumn{4}{c}{Starburst, solar-N/O best fit} & \multicolumn{4}{c}{Starburst, O/H-scaled-N/O best fit}\\
\colhead{} & \colhead{OII/H$\beta$} & \colhead{OIII/H$\beta$} &  \colhead{NII/H$\beta$} & \colhead{SII/H$\beta$} & \colhead{OII/H$\beta$} & \colhead{OIII/H$\beta$} &  \colhead{NII/H$\beta$} & \colhead{SII/H$\beta$} & \colhead{OII/H$\beta$} & \colhead{OIII/H$\beta$} &  \colhead{NII/H$\beta$} & \colhead{SII/H$\beta$} & \colhead{OII/H$\beta$} & \colhead{OIII/H$\beta$} &  \colhead{NII/H$\beta$} & \colhead{SII/H$\beta$} & \colhead{OII/H$\beta$} & \colhead{OIII/H$\beta$} &  \colhead{NII/H$\beta$} & \colhead{SII/H$\beta$}\\
\colhead{(1)} & \colhead{(2)} & \colhead{(3)} & \colhead{(4)} & \colhead{(5)} &\colhead{(6)} & \colhead{(7)} & \colhead{(8)} & \colhead{(9)} & \colhead{(10)} & \colhead{(11)} & \colhead{(12)} &\colhead{(13)} & \colhead{(14)} & \colhead{(15)} & \colhead{(16)} & \colhead{(17)} & \colhead{(18)} & \colhead{(19)} &\colhead{(20)} & \colhead{(21)} 
}
\startdata
\cutinhead{M\,83 (NGC~5236)}
NGC~5236: XUV 01 &$+$0.110 &$+$0.490 &$-$0.434 &$-$0.654 &$+$0.116 &$+$0.410 &$-$0.484 &$-$0.509 &$-$0.172 &$+$0.476 &$-$0.623 &$-$0.209 &$+$0.122 &$+$0.362 &$-$0.521 &$-$1.100 &$+$0.116 &$+$0.397 &$-$0.738 &$-$0.968\\
NGC~5236: XUV 07 &$+$0.660 &$-$0.070 &$-$0.144 &$-$0.524 &$+$0.291 &$+$0.019 &$-$0.215 &$-$0.377 &$+$0.259 &$-$0.182 &$-$0.449 &$-$0.216 &$+$0.327 &$-$0.179 &$-$0.205 &$-$0.895 &$+$0.339 &$-$0.550 &$-$0.612 &$-$1.182\\
NGC~5236: XUV 09 &$+$0.330 &$+$0.390 &$-$0.434 &$-$0.424 &$+$0.146 &$+$0.393 &$-$0.422 &$-$0.407 &$+$0.011 &$+$0.285 &$-$0.606 &$-$0.077 &$+$0.198 &$+$0.263 &$-$0.414 &$-$0.970 &$+$0.116 &$+$0.397 &$-$0.738 &$-$0.968\\
NGC~5236: XUV 10 &$+$0.310 &$+$0.030 &$-$0.354 &$-$0.304 &$+$0.216 &$-$0.067 &$-$0.408 &$-$0.338 &$+$0.364 &$-$0.028 &$-$0.520 &$-$0.182 &$+$0.240 &$+$0.176 &$-$0.348 &$-$0.844 &$-$0.041 &$-$0.027 &$-$0.739 &$-$1.013\\
NGC~5236: XUV 11 &$+$0.310 &$+$0.480 &$-$0.564 &$-$0.654 &$+$0.116 &$+$0.410 &$-$0.484 &$-$0.509 &$+$0.043 &$+$0.277 &$-$0.709 &$-$0.456 &$+$0.204 &$+$0.387 &$-$0.751 &$-$0.954 &$+$0.116 &$+$0.397 &$-$0.738 &$-$0.968\\
NGC~5236: XUV 12 &$+$0.180 &$+$0.160 &$-$0.064 &$-$0.384 &$+$0.121 &$+$0.232 &$-$0.148 &$-$0.499 &$+$0.364 &$-$0.028 &$-$0.520 &$-$0.182 &$+$0.241 &$+$0.177 &$-$0.348 &$-$0.844 &$+$0.116 &$+$0.397 &$-$0.738 &$-$0.968\\
NGC~5236: XUV 13 &$+$0.280 &$+$0.320 &$-$0.234 &$-$0.554 &$+$0.209 &$+$0.249 &$-$0.202 &$-$0.566 &$-$0.138 &$+$0.201 &$-$0.590 &$-$0.494 &$+$0.241 &$+$0.178 &$-$0.348 &$-$0.843 &$+$0.116 &$+$0.397 &$-$0.738 &$-$0.968\\
NGC~5236: XUV 14 &$+$0.160 &$-$0.130 &$-$0.004 &$-$0.284 &$+$0.261 &$-$0.084 &$+$0.060 &$-$0.237 &$+$0.018 &$-$0.390 &$-$0.347 &$-$0.387 &$+$0.223 &$-$0.277 &$-$0.186 &$-$0.699 &$+$0.192 &$-$0.771 &$-$0.058 &$-$1.202\\
NGC~5236: XUV 17 &$+$0.630 &$+$0.050 &$-$0.394 &$-$0.194 &$+$0.307 &$+$0.031 &$-$0.230 &$-$0.156 &$+$0.364 &$-$0.028 &$-$0.520 &$-$0.182 &$+$0.241 &$+$0.177 &$-$0.348 &$-$0.844 &$+$0.147 &$+$0.077 &$-$0.865 &$-$0.918\\
NGC~5236: XUV 20 &$-$0.050 &$-$0.540 &$+$0.076 &\nodata  &$+$0.104 &$-$0.535 &$+$0.032 &\nodata  &$-$0.201 &$-$0.478 &$-$0.309 &\nodata  &$+$0.154 &$-$0.627 &$-$0.135 &\nodata  &$+$0.192 &$-$0.771 &$-$0.058 &\nodata \\
NGC~5236: XUV 21 &$-$0.300 &$-$0.590 &$+$0.076 &$-$0.474 &$-$0.092 &$-$0.616 &$-$0.018 &$-$0.418 &$-$0.204 &$-$0.679 &$-$0.300 &$-$0.457 &$+$0.154 &$-$0.627 &$-$0.135 &$-$0.801 &$+$0.192 &$-$0.771 &$-$0.058 &$-$1.202\\
NGC~5236: XUV 22 &$-$0.290 &$-$0.270 &$+$0.076 &$-$0.323 &$-$0.074 &$-$0.426 &$-$0.028 &$-$0.440 &$-$0.194 &$-$0.456 &$-$0.310 &$-$0.468 &$+$0.102 &$-$0.362 &$-$0.213 &$-$0.700 &$+$0.192 &$-$0.771 &$-$0.058 &$-$1.202\\
\cutinhead{NGC~4625}
NGC~4625: XUV 01 &$+$0.080 &$+$0.590 &$-$0.804 &\nodata  &$-$0.096 &$+$0.506 &$-$0.787 &\nodata  &$+$0.141 &$+$0.547 &$-$0.802 &\nodata  &$+$0.240 &$+$0.654 &$-$0.806 &\nodata  &$+$0.197 &$+$0.690 &$-$0.748 &\nodata \\
NGC~4625: XUV 02 &$+$0.730 &$+$0.300 &$-$0.594 &$-$0.484 &$+$0.128 &$+$0.271 &$-$0.522 &$-$0.372 &$+$0.386 &$+$0.333 &$-$0.709 &$+$0.141 &$+$0.204 &$+$0.387 &$-$0.751 &$-$0.954 &$+$0.116 &$+$0.397 &$-$0.738 &$-$0.968\\
NGC~4625: XUV 05 &$+$0.470 &$+$0.220 &$-$0.354 &$-$0.333 &$+$0.240 &$+$0.239 &$-$0.336 &$-$0.312 &$+$0.364 &$-$0.028 &$-$0.520 &$-$0.182 &$+$0.241 &$+$0.178 &$-$0.348 &$-$0.843 &$+$0.116 &$+$0.397 &$-$0.738 &$-$0.968\\
NGC~4625: XUV 06 &$+$0.580 &$-$0.070 &$-$0.354 &$-$0.203 &$+$0.307 &$+$0.031 &$-$0.230 &$-$0.156 &$+$0.364 &$-$0.028 &$-$0.520 &$-$0.182 &$+$0.277 &$-$0.146 &$-$0.217 &$-$0.838 &$-$0.041 &$-$0.027 &$-$0.739 &$-$1.013\\
NGC~4625: XUV 07 &$+$0.530 &$+$0.440 &$-$0.634 &$-$0.583 &$+$0.119 &$+$0.406 &$-$0.483 &$-$0.510 &$+$0.361 &$+$0.528 &$-$0.754 &$-$0.100 &$+$0.204 &$+$0.387 &$-$0.751 &$-$0.954 &$+$0.116 &$+$0.397 &$-$0.738 &$-$0.968\\
NGC~4625: XUV 09 &$+$0.710 &$+$0.960 &$-$0.394 &\nodata  &$+$0.162 &$+$0.814 &$-$0.385 &\nodata  &$+$0.722 &$+$0.888 &$-$0.337 &\nodata  &$+$0.223 &$+$0.918 &$-$0.783 &\nodata  &$+$0.221 &$+$0.917 &$-$0.735 &\nodata \\
NGC~4625: XUV 12 &$+$0.620 &$+$0.070 &$-$0.644 &$-$0.234 &$+$0.157 &$+$0.116 &$-$0.468 &$-$0.410 &$+$0.364 &$-$0.028 &$-$0.520 &$-$0.182 &$+$0.105 &$+$0.038 &$-$0.690 &$-$0.947 &$+$0.147 &$+$0.077 &$-$0.865 &$-$0.918\\
\enddata
\tablecomments{(1) Region identification, (2) logarithm of the measured [OII]$\lambda\lambda$3726,3729\AA\AA/H$\beta$ line ratio (corrected for extinction), (3) logarithm of the measured [OIII]$\lambda$5007\AA/H$\beta$ line ratio, (4) logarithm of the measured [NII]$\lambda$6584\AA/H$\beta$ line ratio, (5) logarithm of the measured [SII]$\lambda\lambda$6717,6731\AA\AA/H$\beta$ line ratio, (6-9) the same as (2-5) for the line ratios predicted by the best-fitting photoionization model assuming a solar N/O ratio and a single star as ionization source, (10-13) the same as (6-9) for models with O/H-scaled N/O ratio, (14-17) the same as (6-9) for models with solar N/O ratio but a starburst as ionization source, (18-21) the same as (6-9) for models with O/H-scale N/O ratio and a starburst as ionization source. 
}
\end{deluxetable}

\begin{figure}
\figurenum{1}
\epsscale{1.0}
\plotone{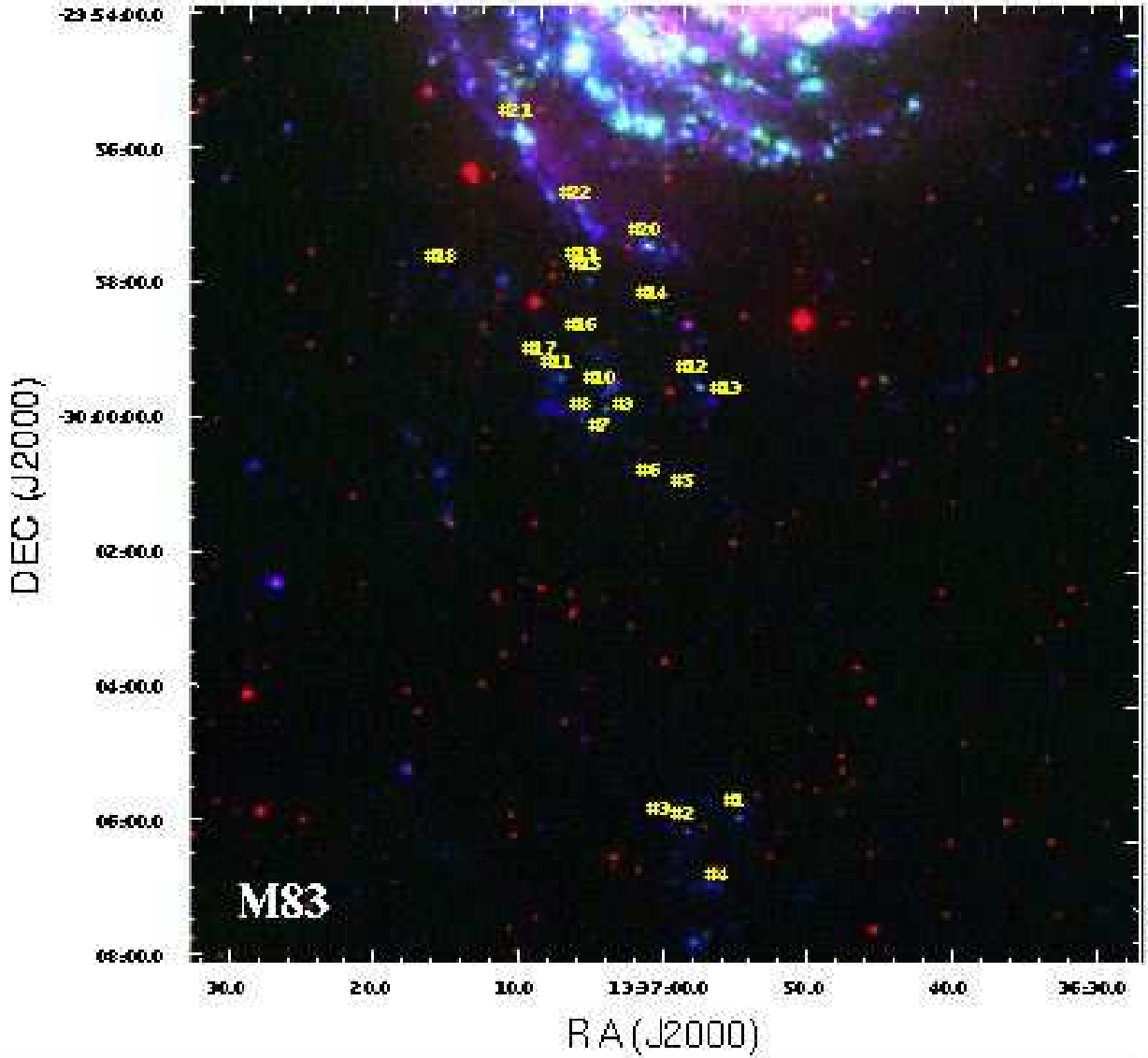}
\caption{False-color RGB image of the southern region of the XUV disk of M\,83. This panel is a composition of the GALEX FUV (blue), ground-based continuum-subtracted H$\alpha$ (green), and $U$-band (red) images. The H$\alpha$-selected regions whose spectroscopy we present in this paper appear as green compact sources in this figure and are located right below the corresponding region identification number (see Table~2).\label{fig1}}
\epsscale{1.0}
\end{figure}

\begin{figure}
\figurenum{2}
\hspace{-2cm}\resizebox{1.1\hsize}{!}{\includegraphics{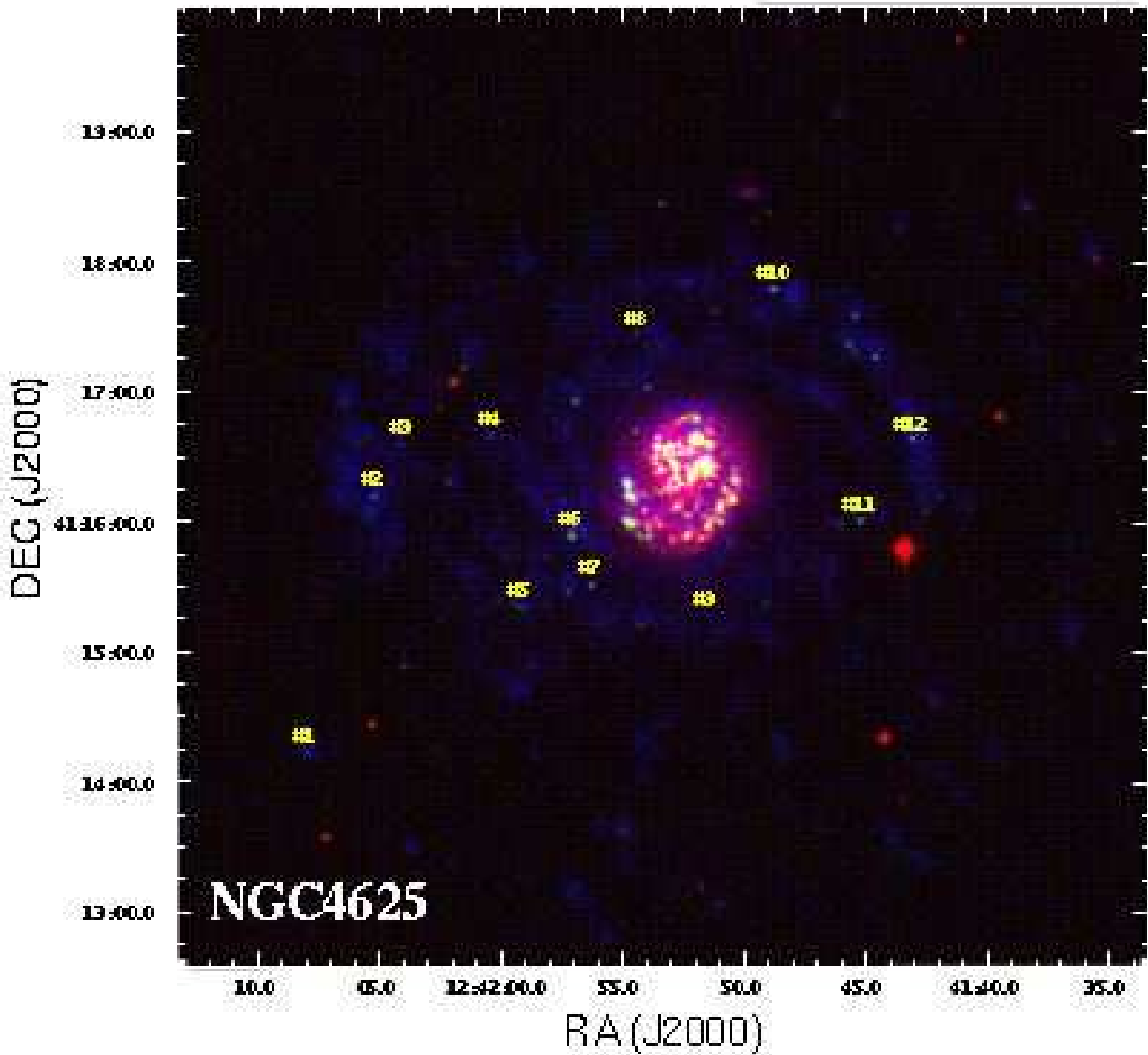}}
\caption{False-color RGB image of the XUV disk of NGC~4625. In this case we show a combination of the smoothed asinh-scaled FUV image (see Gil de Paz et al$.$ 2005a) (blue), the ground-based continuum-subtracted H$\alpha$ image (green), and the $B$-band image (red). The H$\alpha$-selected regions whose spectroscopy we present in this paper appear as green compact sources in this figure and are located right below the corresponding region identification number (see Table~2).\label{fig2}}
\end{figure}

\begin{figure}
\figurenum{3}
\epsscale{.7}
\plotone{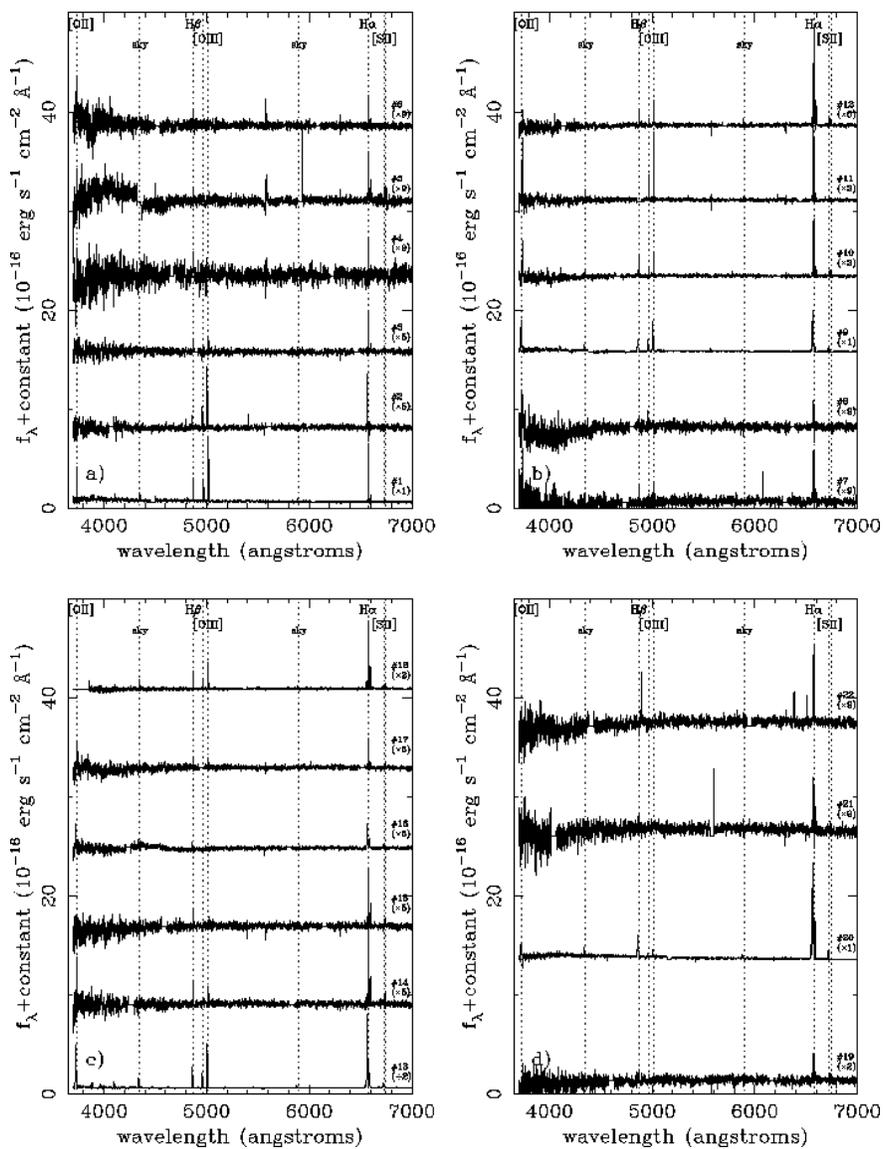}
\caption{Individual spectra of H$\alpha$-selected regions in the XUV disk of M\,83, plus three regions located in its optical disk (NGC~5236: XUV 20, 21, and 22). Only regions with detected H$\alpha$ emission at the approximate redshift of M\,83 are shown. Identification number (according to Table~2) and scaling factors are shown.\label{fig3}}
\epsscale{1.0}
\end{figure}

\begin{figure}
\figurenum{4}
\epsscale{1.0}
\plotone{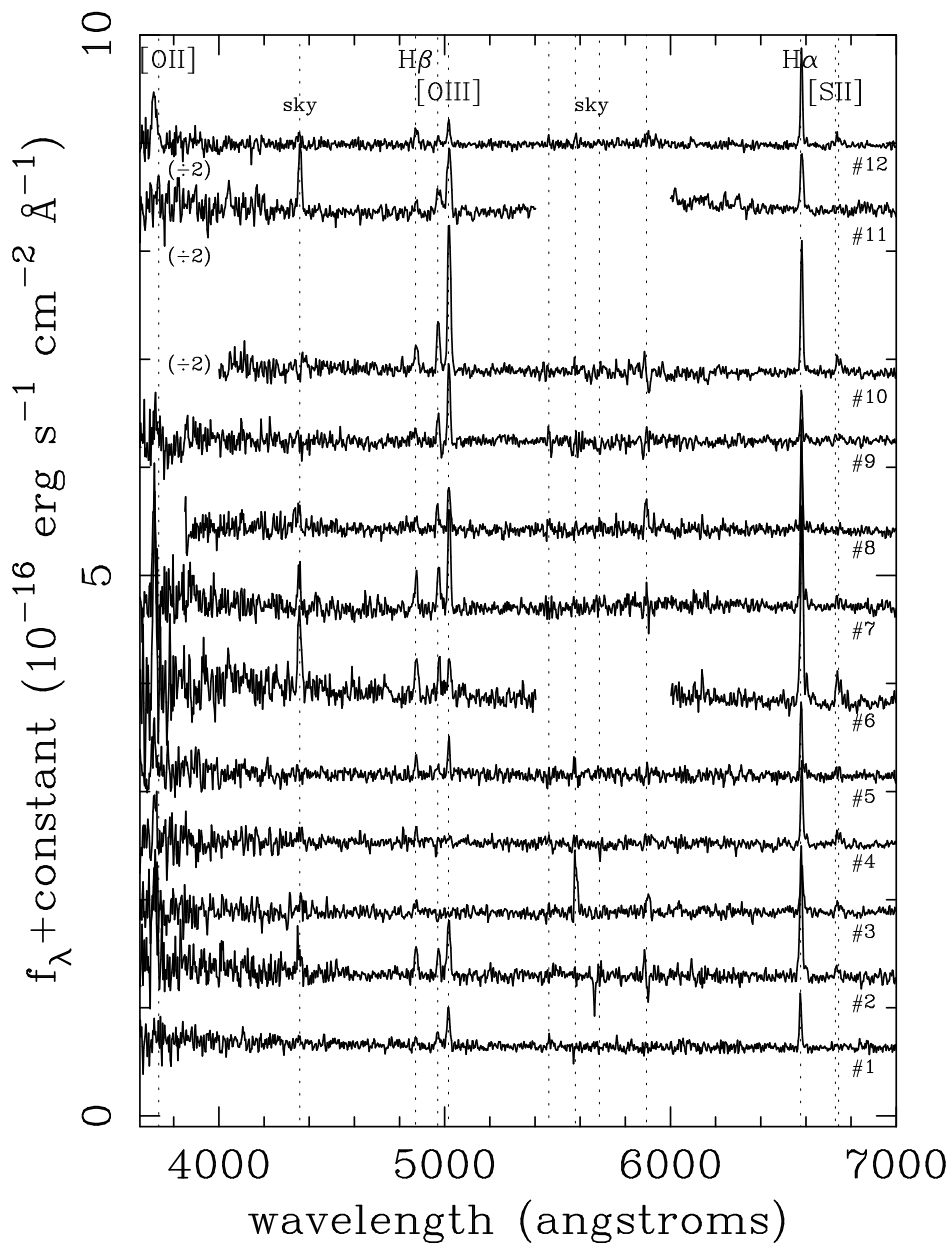}
\caption{Individual spectra of H$\alpha$-selected regions in the XUV disk of NGC~4625. Only regions with detected H$\alpha$ emission at the approximate redshift of NGC~4625 are shown. Identification number (according to Table~2) and scaling factors are shown. Spectral regions with high sky-subtraction residuals in the spectra of NGC~4625: XUV 06 and NGC~4625: XUV 11 have been blanked out for the sake of clarity.\label{fig4}}
\epsscale{1.0}
\end{figure}

\begin{figure}
\figurenum{5}
\epsscale{1.0}
\plotone{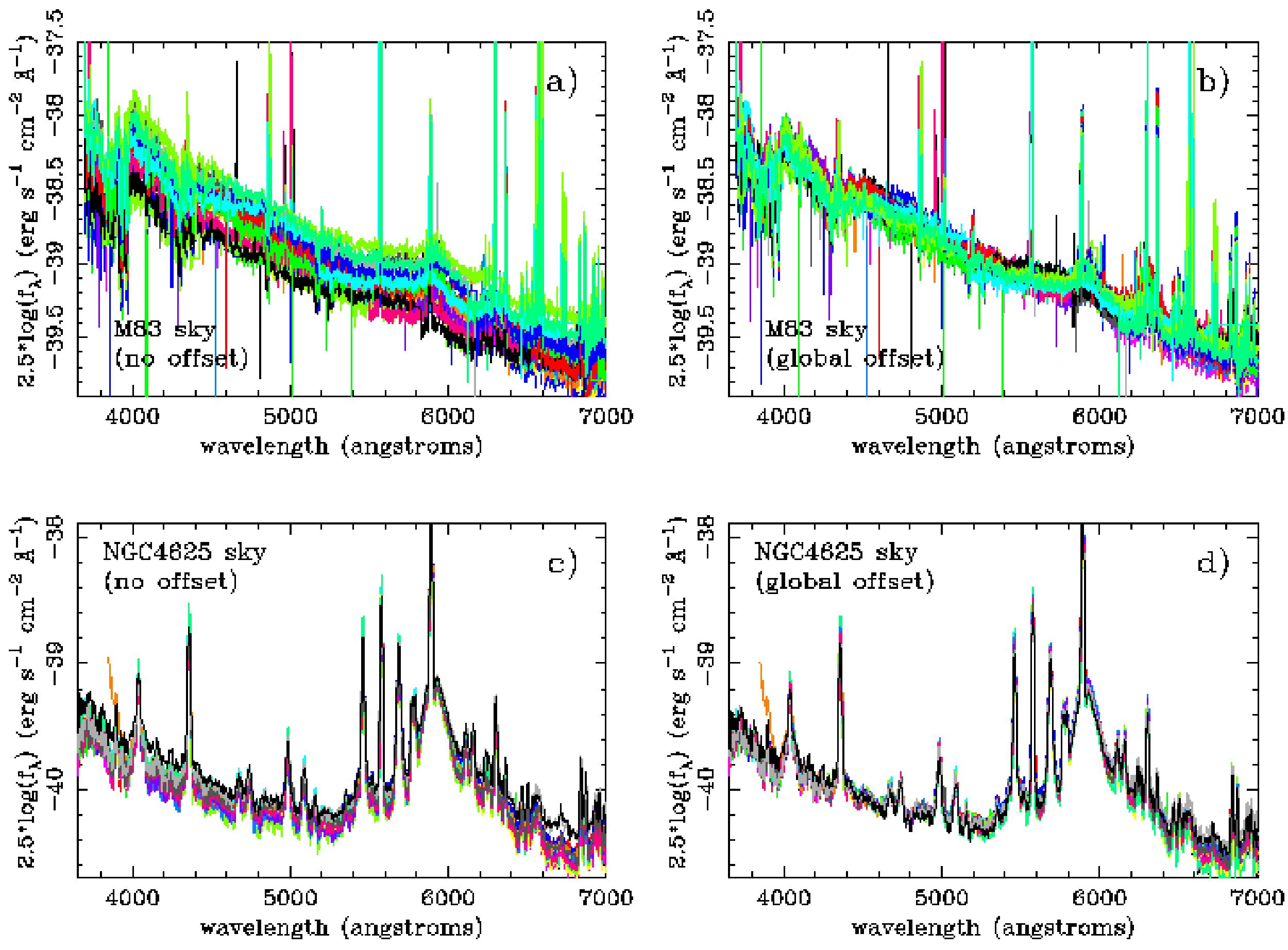}
\caption{{\bf a)} Flux-calibrated sky ($+$object) spectra within each of the slitlets obtained as part of the spectroscopic observations of the XUV disk of M\,83 (different colors are used for clarity). The absolute differences between the individual spectra are of the order of 0.3\,mag. {\bf b)} Differences after small global offsets (the same at all wavelengths but different for each spectrum) are applied. The r.m.s$.$ of the individual spectra are now smaller than 0.2\,mag at almost all wavelengths. {\bf c,d)} The same as {\bf a,b} for the observations of NGC~4625.\label{fig5}}
\epsscale{1.0}
\end{figure}

\clearpage
\begin{center}
\epsscale{1.0}
\plotone{f6a_color.ps}
\plotone{f6b_color.ps}
\epsscale{1.0}
\end{center}
\clearpage
\begin{figure}
\figurenum{6}
\caption{Diagnostic diagrams for the emission-line regions in the XUV disks of M\,83 and NGC~4625. {\bf a)} [OIII]$\lambda$5007\AA/H$\beta$ versus [NII]$\lambda$6584\AA/H$\alpha$ extinction-corrected line ratios. The data for M\,83 are coded based on their galactocentric distance: regions in inner (optical) disk ($r$$<$5.5'; filled circles), regions in the inner XUV disk (5.5'$<$$r$$<$10'; filled triangles), and regions in the outer XUV disk ($r$$>$10'; filled stars). For comparison we show the location of the UCM-Survey galaxies, a complete sample of local star-forming galaxies (SFG) and AGN selected by their emission in H$\alpha$ (Gallego et al$.$ 1996), coded by spectroscopic type. The lines are the predictions of CLOUDY single-star photoionization models for an age of the star of 1\,Myr and a solar N/O abundance ratio (see Section~\ref{cloudy}). Solid lines correspond to models with fixed mass for the ionizing star and metallicities ranging between Z$_{\odot}$/10 and Z$_{\odot}$ while dot-dashed lines correspond to models with fixed metallicity and different masses between 20 and 85\,M$_{\odot}$. Typical errors on the line ratios are shown at the top right corner of the diagram. {\bf b)} [OIII]$\lambda$5007\AA/H$\beta$ versus [OII]$\lambda\lambda$3726,3729\AA\AA/[OIII]$\lambda$5007\AA\ extinction-corrected line ratios. Here we also show the predictions of photoionization models for evolving starbursts (Stasinska \& Leitherer 1996), with different lines showing the predictions for different metallicities (see text for more details on the models). {\bf c)} The same as {\bf a} for NGC~4625. {\bf d)} The same as {\bf b} for NGC~4625. The starburst photoionization models shown here correspond to one-tenth-solar abundance models with two different densities, 10 and 10$^{4}$\,cm$^{-3}$.\label{fig6}}
\epsscale{1.0}
\end{figure}

\begin{figure}
\figurenum{7}
\epsscale{1.0}
\plottwo{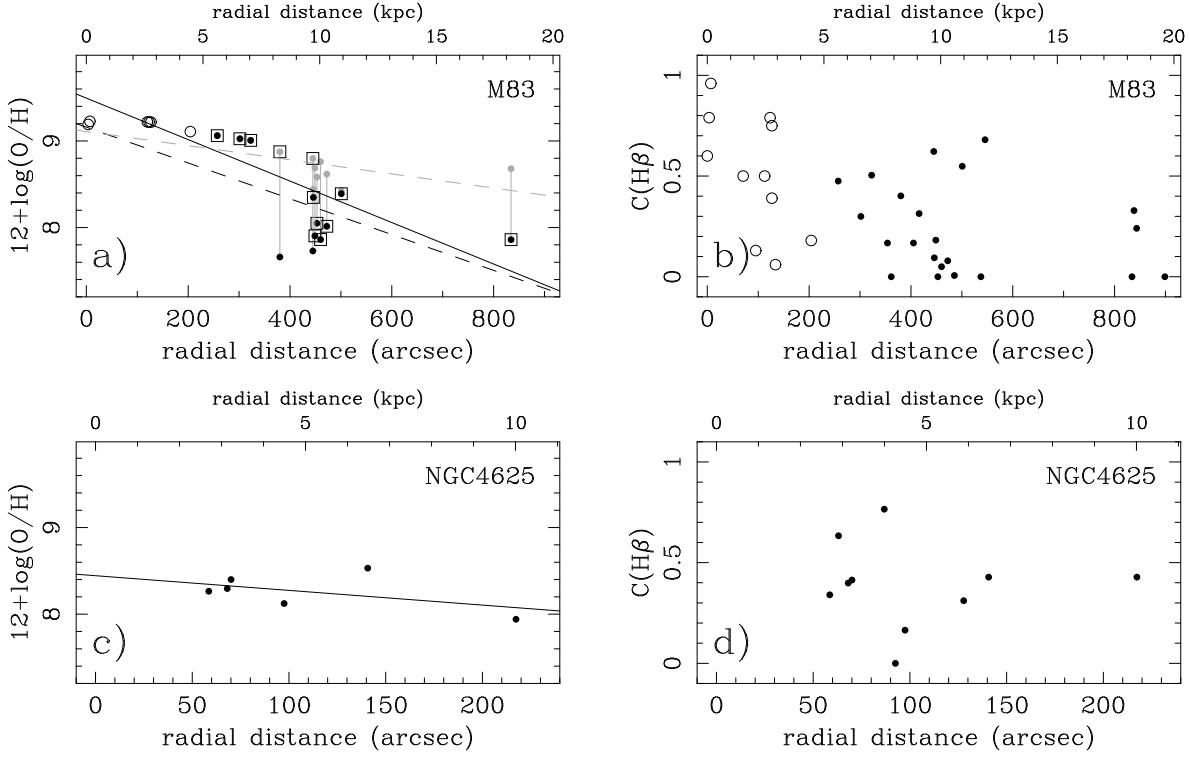}{f7b.ps}
\caption{{\bf a)} Radial distribution of the nebular oxygen abundance in M\,83. Filled dots correspond to metallicities derived from the line ratios measured in this work. Open circles are the metallicities derived using the line ratios measured by published by Webster \& Smith (1983) in six HII regions in the inner optical disk of M\,83. The black (grey) dots are the metallicities obtained assuming the low (high) metallicity branch for the calibration of the R23 parameter for regions NGC~5236: XUV 09 to 14. The black (grey) dashed-line is the corresponding best linear fit. The solid-line is the best fit obtained after using the metal abundance information provided by the best-fitting single-star photoionization models (open squares; see text for details). These oxygen abundances were derived using the McGaugh (1991) calibration of the R23 parameter. Coefficients for all these fits and those obtained using the calibration of Pilyugin \& Thuan (2005) are given in the text. {\bf b)} Radial distribution of the internal extinction in H$\beta$ [C(H$\beta$)] for M\,83. We have adopted the Galactic color excesses in the direction of M\,83 and NGC~4625 to be 0.066 and 0.018\,mag, respectively (Schlegel, Finkbeiner, \& Davis 1998). Symbols have the same meaning as in panel {\bf a}. {\bf c,d)} The same as {\bf a,b} for NGC~4625.\label{fig7}}
\epsscale{1.0}
\end{figure}


\begin{thebibliography}{}
\bibitem[ ]{Al02} Allen, R.J, 2002, in Seeing Through the Dust: The Detection of HI and the Exploration of the ISM in Galaxies, ASP Conference Proceedings, Vol$.$ 276, p$.$ 288. Eds$.$ A.R.Taylor, T.L.Landecker, \& A.G. Willis. San Francisco: Astronomical Society of the Pacific.
\bibitem[ ]{All01} Allende Prieto, C., Lambert, D.L., \& Asplund, M., 2001, ApJ, 556, L63
\bibitem[ ]{Avn76} Avni, Y., 1976, ApJ, 210, 642
\bibitem[ ]{Bal81} Baldwin, J.A., Phillips, M.M., \& Terlevich, R., 1981, PASP, 93, 5
\bibitem[ ]{Bia03} Bianchi, L., Madore, B.F., Thilker, D., Gil de Paz, A., \& Martin, C., 2003, in  The Local Group as an Astrophysical Laboratory, p$.$ 10. Eds$.$ M.Livio \& T.M.Brown. Space Telescope Science Institute.
\bibitem[ ]{Boi00} Boissier, S., \& Prantzos, N., 2000, MNRAS, 312, 398
\bibitem[ ]{Boi03} Boissier, S., Prantzos, N., Boselli, A., \& Gavazzi, G., 2003, MNRAS, 346, 1215  
\bibitem[ ]{Boi06} Boissier, S., et al., 2006, ApJ, in press (astro-ph/0609071)
\bibitem[ ]{Bro06} Brook, C.B., Kawata, D., Martel, H., Gibson, B.K., \& Bailin, J., 2006, ApJ, 639, 126
\bibitem[ ]{Bru03} Bruzual, G., \& Charlot, S., 2003, MNRAS, 344, 1000
\bibitem[ ]{Bus04} Bush, S.J., \& Wilcots, E.M., 2004, AJ, 128, 2789
\bibitem[ ]{Car89} Cardelli, J.A., Clayton, G.C., \& Mathis, J.S., 1989, ApJ, 345, 245
\bibitem[ ]{Cer98} Cervi\~no, M., 1998, Ph.D$.$ Thesis, Universidad Complutense de Madrid, Spain.
\bibitem[ ]{Cla89} Clarke, C.J., 1989, MNRAS, 238, 283
\bibitem[ ]{Dra84} Draine, B.T., \& Lee, H.M., 1984, ApJ, 285, 89
\bibitem[ ]{Elm06} Elmegreen, B.G., \& Hunter, D.A., 2006, ApJ, 636, 712
\bibitem[ ]{Erw05} Erwin, P., Beckman, J.E., \& Pohlen, M., 2005, ApJ, 626, L81
\bibitem[ ]{Frg97} Ferguson, A.M.N., 1997, Ph.D. thesis, Johns Hopkins Univ. 
\bibitem[ ]{Frg98} Ferguson, A.M.N., Gallagher, J.S., \& Wyse, R.F.G., 1998, AJ, 116, 673
\bibitem[ ]{Fer98} Ferland, G.J., Korista, K.T., Verner, D.A., Ferguson, J.W., Kingdon, J.B., \& Verner, E.M., 1998, PASP, 110, 761
\bibitem[ ]{jgm96} Gallego, J., Zamorano, J., Rego, M., Alonso, O., \& Vitores, A.G., 1996, A\&AS, 120, 323
\bibitem[ ]{Gal03} Galliano, F., Madden, S.C., Jones, A.P., Wilson, C.D., Bernard, J.-P., \& Le Peintre, F., 2003, A\&A, 407, 159
\bibitem[ ]{Gal05} Galliano, F., Madden, S.C., Jones, A.P., Wilson, C.D., \& Bernard, J.-P., 2005, A\&A, 434, 867
\bibitem[ ]{BCD03} Gil de Paz, A., Madore, B.F., \& Pevunova, O., 2003, ApJS, 147, 29 
\bibitem[ ]{G05} Gil de Paz, A., et al., 2005, ApJ, 627, L29
\bibitem[ ]{2003IAUS..212...70H} Hillier, D.J., 2003, IAU Symposium, 212, 70
\bibitem[ ]{HM79} Hollenbach, D., \& McKee, C.F., 1979, ApJS, 41, 555 
\bibitem[ ]{HS71} Hollenbach, D., \& Salpeter, E.E., 1971, ApJ, 163, 155
\bibitem[ ]{Hoo05} Hoopes, C.G., et al., 2005, ApJ, 619, L99
\bibitem[ ]{Hou00} Hou, J.L., Prantzos, N., \& Boissier, S., 2000, A\&A, 362, 921
\bibitem[ ]{Hug71} Hughes, M.P., Thompson, A.R., \& Colvin, R.S., 1971, ApJS, 23, 323
\bibitem[ ]{Jon79} Jones, T.W., \& Merrill, K.M., 1976, ApJ, 209, 509 
\bibitem[ ]{Keh06} Kehrig, C., V\'{\i}lchez, J.M., Telles, E., Cuisinier, F., \& P\'{e}rez-Montero, E., 2006, A\&A, 457, 477 
\bibitem[ ]{Kel98} Kells, W., et al., 1998, PASP, 110, 1487
\bibitem[ ]{Kew02} Kewley, L.J., \& Dopita, M.A., 2002, ApJS, 142, 35
\bibitem[ ]{Ken89} Kennicutt, R.C.Jr., 1989, ApJ, 344, 685  
\bibitem[ ]{Kn74} Knapp, G.R., \& Kerr, F.J., 1974, A\&A, 35, 361
\bibitem[ ]{Koh01} Kohta, N., Naomasa, N., \& Kuno, N., 2001, PASJ, 53, 757
\bibitem[ ]{Kop05} K\"oppen, J., \& Hensler, G., 2005, A\&A, 434, 531 
\bibitem[ ]{Kro93} Kroupa, P., Tout, C.A., \& Gilmore, G., 1993, MNRAS, 262, 545
\bibitem[ ]{Lar76} Larson, R.B., 1976, MNRAS, 176, 31
\bibitem[ ]{Sb99} Leitherer, C., et al., 1999, ApJS, 123, 3
\bibitem[ ]{Lej97} Lejeune, T., Cuisinier, F., \& Buser, R., 1997, A\&AS, 125, 229
\bibitem[ ]{Liang06} Liang, Y.C., Yin, S.Y., Hammer, F., Deng, L.C., Flores, H., \& Zhang, B., 2006, ApJ, 652, 257
\bibitem[ ]{Mad04} Madore, B.F., et al., 2004, American Astronomical Society Meeting 205, \#128.01  
\bibitem[ ]{Mar01} Martin, C.L., \& Kennicutt, R.C.Jr., 2001, ApJ, 555, 301
\bibitem[ ]{Mar05} Martin, D.C., et al., 2005, ApJ, 619, L1
\bibitem[ ]{FM89} Matteucci, F., \& Francois, P., 1989, MNRAS, 239, 885
\bibitem[ ]{McG91} McGaugh, S.S., 1991, ApJ, 380, 140 (M91)
\bibitem[ ]{Meu04} Meurer, G.R., Thilker, D., Bianchi, L., Ferguson, A., Madore, B.F., \& Gil de Paz, A., 2004, American Astronomical Society Meeting 205, \#42.03
\bibitem[ ]{Mih99} Mihos, J.C., Spaans, M., \& McGaugh, S.S., 1999, ApJ, 515, 89
\bibitem[ ]{Moe06} Moehler, S., \& Sweigart, A.V., 2006, A\&A, 455, 943
\bibitem[ ]{Molla} Moll\'{a}, M., V\'{\i}lchez, J.M., Gavil\'{a}n, M., \& D\'{\i}az,A.I., 2006, MNRAS, 372, 1069 
\bibitem[ ]{jcm06} Mu\~noz-Mateos, J.C., Gil de Paz, A., Boissier, S., Zamorano, J., Jarrett, T., Gallego, J., \& Madore, B.F., 2006, ApJ, in press (astro-ph/0612017). 
\bibitem[ ]{O89} Osterbrock, D.E., 1989, Astrophysics of Gaseous Nebulae and Active Galactic Nuclei. Mill Valley: University Science Books.
\bibitem[ ]{2005AIPC..804..105P} Pauldrach, A.W.A., 2005, AIP Conf.~Proc.~804: Planetary Nebulae as Astronomical Tools, 804, 105 
\bibitem[ ]{pt05} Pilyugin, L.S., \& Thuan, T.X., 2005, ApJ, 631, 231 (PT05)
\bibitem[ ]{pt00} Pilyugin, L.S., 2000, A\&A, 362, 325
\bibitem[ ]{Pop03} Popescu, C.C., \& Tuffs, R.J., 2003, A\&A, 410, L21
\bibitem[ ]{P00} Prantzos, N., \& Boissier, S., 2000, MNRAS, 313, 338
\bibitem[ ]{Sch97} Schaerer, D., \& de Koter, A., 1997, A\&A, 322, 598
\bibitem[ ]{Sch98} Schlegel, D.J., Finkbeiner, D.P., \& Davis, M., 1998, ApJ, 500, 525
\bibitem[ ]{Smi02} Smith, L.J., Norris, R.P.F., \& Crowther, P.A., 2002, MNRAS, 337, 1309
\bibitem[ ]{Spi78} Spitzer, L., 1978, Physical processes in the interstellar medium. New York Wiley-Interscience.
\bibitem[ ]{Sta96} Stasinska, G., \& Leitherer, C., 1996, ApJS, 107, 661
\bibitem[ ]{Sta97} Stasinska, G., \& Schaerer, D., 1997, A\&A, 322, 615
\bibitem[ ]{Sw02} Swaters, R.A., \& Balcells, M., 2002, A\&A, 390, 863 
\bibitem[ ]{T05a} Thilker, D.A., et al., 2005a, ApJ, 619, L79
\bibitem[ ]{T05b} Thilker, D.A., et al., 2005b, American Astronomical Society Meeting 207, \#202.02 
\bibitem[ ]{Til93} Tilanus, R.P.J., \& Allen, R.J., 1993, A\&A, 274, 707
\bibitem[ ]{Tout} Tout, C.A., Pols, O.R., Eggleton, P.P., \& Han, Z., 1996, MNRAS, 281, 257
\bibitem[ ]{Tru05} Trujillo, I., \& Pohlen, M., 2005, ApJ, 630, L17
\bibitem[ ]{van06} van Zee, L., \& Haynes, M.P., 2006, ApJ, 636, 214
\bibitem[ ]{van98a} van Zee, L., Salzer, J.J., \& Haynes, M.P., 1998a, ApJ, 497, L1 
\bibitem[ ]{van98b} van Zee, L., Salzer, J.J., Haynes, M.P., O'Donoghue, A.A., \& Balonek, T.J., 1998b, AJ, 116, 2805  
\bibitem[ ]{Vid05} Vidali, G., et al., 2005, in Light, Dust and Chemical Evolution, Journal of Physics: Conference Series, Vol$.$ 6, 36. Institute of Physics. 
\bibitem[ ]{Vei87} Veilleux, S., \& Osterbrock, D.E., 1987, ApJS, 63, 295
\bibitem[ ]{Web83} Webster, B.L., \& Smith, M.G., 1983, MNRAS, 204, 743
\bibitem[ ]{Zam04} Zamorano, J., Gallego, J., Rego, M., Vitores, A.G., \& Alonso, O., 1996, ApJS, 105, 343
\bibitem[ ]{Zam06} Zamorano, J., Rego, M., Gallego, J., Vitores, A.G., Gonz\'{a}lez-Riestra, R., \& Rodr\'{\i}guez-Caderot, G., 1994, ApJS, 95, 387
\end{thebibliography}
\end{document}